\documentclass[]{aastex631}

\usepackage{graphicx} 
\usepackage{booktabs} 
\usepackage{subfigure}
\usepackage{overpic}    
\usepackage{xcolor}  
\usepackage{float}     
\usepackage{listings} 
\usepackage{xcolor}   
\usepackage{amsmath}  
\usepackage{cleveref}
\usepackage{bm}

\lstdefinestyle{PythonStyle}{
    language=Python,
    basicstyle=\ttfamily\footnotesize,
    keywordstyle=\color{blue},
    commentstyle=\color{gray},
    stringstyle=\color{red},
    breaklines=true}

\shorttitle{The Milky Way Bar}
\shortauthors{Kumar et al.}
\graphicspath{{./}{figures/}}

\begin{document}

\title{The Expanding 3 kpc Arms Are Neither Expanding nor Spiral Arms but X1 Orbits Driven by the Galactic Bar}

\author[0000-0002-9571-8036]{Jayender Kumar}
\affiliation{CSIRO Space and Astronomy, ATNF Headquarters \\
26 Pembroke Rd, Marsfield, NSW 2122, Australia}
\affiliation{School of Natural Sciences, University of Tasmania, \\
Private Bag 37, Hobart, Tasmania 7001, Australia}

\author[0000-0001-7223-754X]{Mark J. Reid}
\affiliation{Center for Astrophysics $\vert$ Harvard \& Smithsonian, \\
60 Garden Street, Cambridge, Massachusetts 02138, USA}

\author[0000-0003-0109-2392]{T. M. Dame}
\affiliation{Center for Astrophysics $\vert$ Harvard \& Smithsonian, \\
60 Garden Street, Cambridge, Massachusetts 02138, USA}

\author[0000-0002-1363-5457]{Simon P. Ellingsen}
\affiliation{School of Natural Sciences, University of Tasmania, \\
Private Bag 37, Hobart, Tasmania 7001, Australia}
\affiliation{International Centre for Radio Astronomy Research, The University of Western Australia, \\
35 Stirling Highway, Crawley WA 6009, Australia}

\author[0000-0002-4783-6679]{Lucas J. Hyland}
\affiliation{School of Natural Sciences, University of Tasmania, \\
Private Bag 37, Hobart, Tasmania 7001, Australia}

\author[0000-0003-4468-761X]{Andreas Brunthaler}
\affiliation{Max-Planck-Institut für Radioastronomie, \\
Auf dem Hügel 69, Bonn 53121, Germany}

\author[0000-0001-6459-0669]{Karl M. Menten}
\affiliation{Max-Planck-Institut für Radioastronomie, \\
Auf dem Hügel 69, Bonn 53121, Germany}

\author{Xing-Wu Zheng}
\affiliation{Department of Astronomy, Nanjing University, \\
Hankou Road, Nanjing 210093, People's Republic of China}

\author[0000-0001-7960-4912]{Alberto Sanna}
\affiliation{INAF, Osservatorio Astronomico di Cagliari, \\
via della Scienza 5, Selargius(CA) 09047, Italy}

\begin{abstract}
Near the center of our Milky Way is a bar-like structure and the so-called Expanding 3 kpc arms. We currently have limited knowledge of this important region, since we are about 8.2 kpc from the center and cannot directly observe it at optical wavelengths, owing to strong extinction from interstellar dust. Here we present extremely precise very long baseline interferometry measurements of H$_2$O  maser sources from the BeSSeL Survey, where extinction is not a problem, which accurately determine the three-dimensional locations and motions of three massive young stars. Combined with previous measurements, these stars delineate a trail of orbits outlining the Milky Way’s Galactic bar. We present the first measurements capturing the dynamics of quasi-elliptical (X1) orbits around the Galactic bar. Our findings provide evidence substantiating the existence of such orbits populated by massive young stars. Our measurements of the position and velocity of a number of massive young stars, previously identified with the Expanding 3 kpc arms, show that they are more likely located in the X1 orbits about the Galactic bar. Also, some stars previously assigned to the Norma spiral arm appear to be in these orbits, which suggests that this spiral arm does not extend past the end of the bar. 

\end{abstract}



\section{Introduction}\label{sec:intro}

The Milky Way, the spiral galaxy that we live in, is a vast body with a visible size of roughly 40 kpc. It is made up of copious amounts of stars, dust, and gas. Our solar system is at a distance of 8.2 kpc from the Galactic center \citep{Reid:2019, Do:2019, GRAVITYCollaboration:2021}. Most of the Milky Way, especially its inner region and its far side, is hidden from our sight due to optical extinction through the Galactic plane. The Milky Way’s disk contains arms that spiral outward from a bar-like structure. This Galactic bar extends about 4.5 kpc on each side of the Galactic center and is tilted by about $30^\circ$ from our line of sight to the center \citep{Wegg:2015}. 

The 3 kpc arm was originally discovered by \citet*{vanWoerden:1957} as an absorption feature in hydrogen at Galactic longitude 0$^{\circ}$, with radial velocities around $-53$~km s$^{-1}$. Oort gave it its original name of ``the Expanding 3 kpc arm'', as gas in the arm appeared to be expanding outwards from the Galactic center at this velocity \citep{Oort:1958, Oort:1977}. Since the arm was located on the near side of the Galactic center, it was also referred to as the ``Near 3 kpc arm'' in later studies\footnote{Hereafter, we refer to \cite{Oort:1958}'s ``Expanding 3 kpc arm'' as the ``Near 3 kpc'' arm.}. Recently, its symmetric counterpart on the far side of the Galactic center, called the ``Far 3 kpc arm'' was finally detected \citep{Dame:2008}. The arms were primarily traced through radio and millimeter observations due to the heavy dust obscuration in the Galactic plane.

\begin{figure}[h]%
\centering
\includegraphics[width=0.7\textwidth]{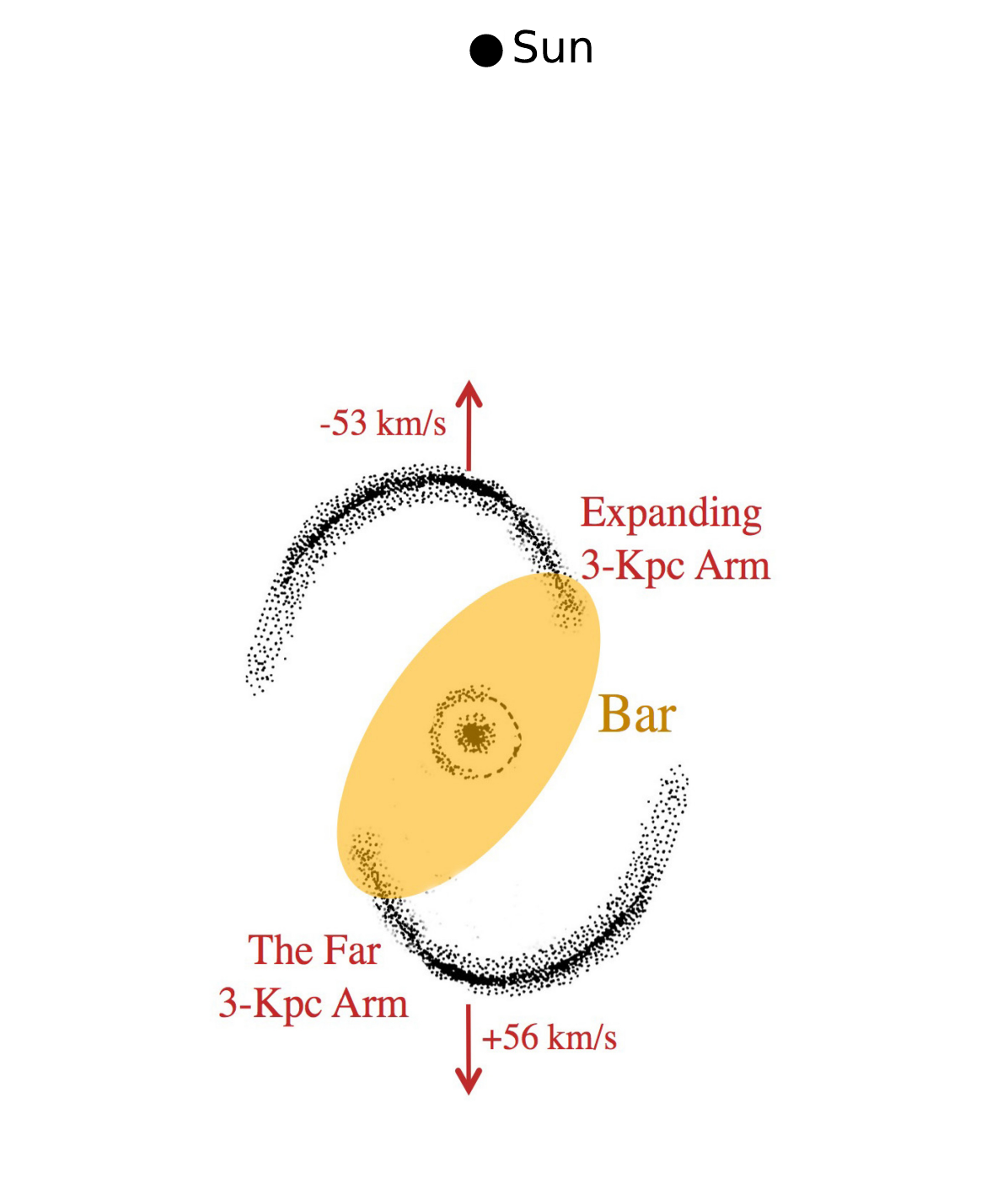}
\caption{A model of the Expanding 3 kpc spiral arms adapted from Figure 9 of \cite{Oort:1977} (see also \url{https://www.cfa.harvard.edu/news/milky-ways-inner-beauty-revealed}; image credit: Thomas M. Dame). The model is based on hydrogen emission from near the Galactic bar (gold ellipse) at the center of the Milky Way. The Expanding Near and Far 3 kpc arms are shown as black spiral traces. The Sun is at the top. The red arrows attached to the arms indicate observed motions of gas along our line of sight toward the Galactic center, leading to the interpretation of expansion from the center \citep{Oort:1977, RougoorOort:1960}.
}
\label{3kpctoymodel}
\end{figure}

Since their identification, the Expanding 3 kpc arms have remained amongst the most puzzling structures in the Milky Way. Both the near and far parts were found to be moving with velocities at around $55$~km s$^{-1}$, with the near part approaching us and the far part receding.
Figure~\ref{3kpctoymodel} gives a schematic representation of these structures. 

\cite{Dame:2008} using CO emission in a Galactic longitude versus line-of-sight velocity ($l-v$) diagram identified and traced both parts of the Expanding 3 kpc arms within Galactic longitudes of $\pm12^{\circ}$. At larger longitude and velocity offsets, they are hard to distinguish among different features in an $l-v$ diagram.

Soon after the discovery of the Near 3 kpc arm, alternative theories were suggested to explain this feature in the inner Milky Way. Initially, the large noncircular expansion of the arms was thought to result from the explosion of gas from the Galactic center \citep{vanderKruit:1971, SandersPrendergast:1974}. \cite{CohenDavies:1976}, using HI emission data for these arms, fitted a Galactocentric ring model of gas with a radius of 4 kpc, which rotates at 210~km s$^{-1}$ and expands at between 53 and 56~km s$^{-1}$. \cite{Bania:1980} reported an arm tangency at $+$23.5$^{\circ}$ longitude by superimposing the \cite{CohenDavies:1976} ring model on their CO emission data.

\cite{Peters:1975} offered an alternative explanation for the apparent expansion of the 3 kpc arms, which involved the gas moving in elliptical, rather than circular, orbits. \cite{ContopoulosPapayannopoulos:1980} suggested that gas in the inner region of the Galaxy moved in resonant (X1 and X2) orbits about a bar. Over the years, observations at infrared wavelengths suggested that the Galactic bar was oriented with its long axis tilted at an azimuthal angle from our line of sight to the center of $30^\circ$ to $45^\circ$ \citep{Hammersley:1994, Hammersley:2000, Benjamin:2005, Zasowski:2012, Wegg:2015}.

While the expanding nature of the 3 kpc arms has been quietly sidelined over the intervening decades, no strong evidence has been forwarded for this and the true nature of these arms has remained an unsolved problem. Lacking accurate distance measurements, most efforts to trace the 3 kpc arms have used $l-v$ diagrams of CO and HI emission and absorption features. Unfortunately, in this crowded region of the Galaxy, many structures can have similar $l-v$ values, which causes confusion. In our work, we approach the task of mapping the inner Milky Way by using radio astrometry as part of the Bar and Spiral Structure Legacy (BeSSeL) Survey \citep{Brunthaler:2011}, which measures trigonometric parallaxes to determine distances to massive young stars\footnote{These stars are extremely young ($\ll1$ Myr) and may not yet have reached the main sequence.} with strong molecular maser emission. We use the Very Long Baseline Array (VLBA) of the National Radio Astronomy Observatory\footnote{The National Radio Astronomy Observatory is a facility of the National Science Foundation operated under cooperative agreement by Associated Universities, Inc.} (NRAO) 
to achieve extremely high angular resolution, enabling position measurements with precision approaching micro-seconds of arc. Importantly, dust extinction, which completely blocks our view of the Galactic Center at optical wavelengths, is insignificant at radio wavelengths. 

Our paper is organized in the following manner. Section~\ref{sec:2} describes our observational strategy, the data calibration procedures, and measurement techniques. In Section~\ref{sec:3}, we present our approach for selecting sources associated with the Galactic bar. Section~\ref{sec:ResultsDiscussion} presents our results, accompanied by an in-depth discussion. Finally, Section~\ref{sec:conclusion} summarizes the conclusions drawn from this work.

\section{New Measurements}\label{sec:2}

We used the VLBA to obtain observations at 16 epochs spanning one year toward three H$_2$O\ maser sources:  G008.83$-$00.02, G010.47$+$00.02, and G024.78$+$00.08. These were conducted under program codes BR210A and BR210B. We observed the $6_{1,6}-5_{2,3}$ H$_2$O\ maser transition at a rest frequency of 22.235080 GHz. Each observation was approximately 7 hours long, with individual maser on-source times near 16 minutes. The dates for each of the epochs are listed in Table~\ref{tab:appB1} in Appendix~\ref{sec:appendixB}, and observational parameters for each source are listed in Table~\ref{tab:appB2} in Appendix~\ref{sec:appendixB}.

\smallskip

We followed standard BeSSeL Survey equipment setups \citep{Reid:2009b}. Observations of the maser targets were interleaved with scans of two extragalactic sources nearby in angle on the sky. The calibrators were selected from a calibrator survey \citep{Immer:2011} and have typical position accuracy of $\pm2$~mas.  Data were recorded in two modes: phase-referencing and geodetic blocks.

\smallskip
In order to calibrate and remove residual atmospheric delays for each antenna, four geodetic blocks \citep{Reid:2009b} were scheduled, one each at the start and end of the observations and two in between. For these observations, we used eight 16-MHz wide intermediate frequency (IF) bands spaced over the frequency range 23.522--24.014~GHz. The lower edge frequencies for each IF were 23.522, 23.536, 23.578, 23.648, 23.732, 23.830, 23.970 and 23.998~GHz. The geodetic block data were recorded only in left circular polarization.

\smallskip
For the phase-referencing blocks, data were recorded from four adjacent 16 MHz IF bands, covering the frequency range 22.1985--22.2625~GHz. Data in both right and left circular polarizations were recorded and cross-correlated in two passes: (1) a continuum mode with spectral channels of 500~kHz and (2) a spectral-line mode with 8~kHz (0.108~km s$^{-1}$) channels covering 216 km s$^{-1}$ in the third IF band containing the maser emission. The raw data were processed using the VLBA DiFX-2 software correlator \citep{Deller:2011} in Socorro, New Mexico. 

\subsection{Astrometric Calibrations}

\smallskip
For data calibration and imaging, we used the NRAO Astronomical Image Processing System with a ParselTongue scripting interface \citep{Kettenis:2006}. We corrected the data for delays due to variations in Earth orientation, changes in parallactic angle (apparent feed rotation), and ionospheric delays using maps of total electron content. Amplitudes in the cross-correlation spectra were then corrected to account for the digitizer sample threshold errors. Measurements of system temperature were used for the amplitude calibration. The electronic delay and phase differences among frequency bands were corrected using a single scan of a strong calibration source. Residual zenith tropospheric delays, clock offsets, and clock drift rates estimated from the geodetic blocks were removed from the data \citep{Reid:2009a}. 

All interferometric spectra for the masers were shifted in frequency to compensate for changes in the Earth's orbital motion throughout the observations. We selected a strong and compact maser channel as the interferometer phase reference. We then transferred the phases from the maser data to the continuum data. The continuum data were then averaged in frequency, imaged, and positions estimated by fitting Gaussian brightness distributions. The measured calibrator position offsets were subtracted from the maser positions. Imaging of the spectral line data used a pixel size of 0.5 mas, an image size of 1024$\times$1024 pixels, and CLEAN restoring beams listed in Table~\ref{tab:appB2} in Appendix~\ref{sec:appendixB}. Figures~\ref{fig:appA1}--\ref{fig:appA3} in Appendix~\ref{sec:appendixA} show interferometer Stokes-I spectra of the maser emission for the three maser sources. These were generated by summing over a 2048$\times$2048 mas region (pixel size = 0.5 mas and an image size of 4096$\times$4096 pixels) from epoch 10 for G008.83$-$00.02 and G010.47$+$00.02 and epoch 9 for G024.78$+$00.08 using the task ISPEC on image cubes.  

\smallskip

\subsection{Parallax and Proper Motion Measurements}
In order to estimate parallax and proper motion, we first examined the emission spectra at selected epochs spanning the observations for maser features that appeared as single (unblended) components with high signal-to-noise ratios. Position offsets for selected maser channels and each calibrator were estimated with the task JMFIT. We subtracted the measured calibrator position offsets from the maser spots. The position offsets of the reference maser spot relative to each calibrator as a function of time are shown in  Figures~\ref{fig:appA4}--\ref{fig:appA6} in Appendix~\ref{sec:appendixA}. We fitted these position offsets to a model that allowed for the parallax effect as well as a linear proper motion. This involves five parameters: a single parallax parameter and position offset and linear motion parameters in the easterly and northerly directions.  "Error floors" were added in quadrature to the formal position uncertainties and adjusted to achieve postfit residuals with a $\chi^2$\ per degree of freedom near unity separately for the easterly and northerly offsets.  Error floors are intended to capture systematic errors typically owing to uncompensated atmospheric delays, as well as possible structural variations in a calibrator. Figures~\ref{fig:appA4}--\ref{fig:appA6} show the parallax fits for the three masers.

\smallskip
In order to improve the robustness of our parallax measurement, for each maser source we also fitted the parallax and proper motion for all maser spots in the emission spectra with a signal-to-noise ratio (peak brightness divided by the rms noise of an image) greater than 10, with detections in at least 10 epochs, and were nearly pointlike and not spatially blended with other spots.  We combined all data relative to one calibrator and fitted for a single parallax, but different proper motions for each spot owing to internal motions among the spots.  In order to optimally weight the data from different spots, we added the error floors estimated from the individual spot fits in quadrature with the measurement errors prior to the combined fit.  Then we multiplied the formal least-squares parallax uncertainty by $\sqrt{N}$, where N is the number of maser spots, in order to account for the possibility of correlated position shifts among the maser spots \citep{Reid:2009a}. 
The final parallax estimate was a variance-weighted average of the parallax values based on the two calibrators, which should be mostly uncorrelated.

\smallskip
In order to estimate the proper motion of the central star that excites a maser source, we employed a hybrid averaging approach: calculating a variance-weighted average of each maser spot relative to both calibrators and then taking an unweighted average of the resultant values for all spots. The parallax and proper motion estimates obtained using this approach for the three sources are summarized in Table~\ref{tab:appB3} in Appendix~\ref{sec:appendixB}. The final results obtained for the parallax and proper motion for each maser source are listed in Table~\ref{tab:appB4} in Appendix~\ref{sec:appendixB}.

\smallskip
G010.47$+$00.02 has a previously reported parallax measurement from the BeSSeL Survey (project code BR145) of $\pi = 0.117\pm0.008$~mas \citep{Sanna:2014}. Our parallax is statistically consistent with their measurement. In order to obtain the best estimate of the parallax, we use a variance-weighted average of the two measurements. This gives a final parallax of $\pi = 0.115\pm0.008$ mas. Similarly, we averaged the proper motions of the two measurements for the best estimates of proper motion. These updated values are listed in Table~\ref{tab:appB4} in Appendix~\ref{sec:appendixB}.

\section{Previously Measured Sources in the Bar Region}\label{sec:3}

In order to define a sample of sources associated with the Galactic bar and with measured parallaxes, we collected 19 published by the BeSSeL Survey \citep{Reid:2019} and the VERA Project \citep{VERACollaboration:2020} that were within 4.2 kpc of the Galactic center. This limit approximates the long bar's semimajor axis, which is estimated in the recent literature to range from 3.4 to 4.6 kpc \citep{GreenCaswell:2011, Wegg:2015}. Adding the three measurements reported here, our initial sample includes 21 sources (G010.47$+$00.02 had a previously published parallax \citep{Sanna:2014}). We excluded two sources that lie within 1 kpc of the center since these should not be associated with the bar. Next, we removed sources with fractional parallax uncertainties greater than $\pm20$\%.  For a characteristic parallax of the Galactic center of 0.122 mas (8.2 kpc), this uncertainty corresponds to a $\pm1\sigma$ distance range of 6.8 to 10.2 kpc, which would preclude locating a source accurately within the bar region. Five more sources were excluded based on this criterion. After applying these selection criteria, we are left with 14 sources. 

Table~\ref{tab:14BarSources} lists coordinates, measured parallaxes, proper motions, and LSR velocities.  Also listed are derived Galactocentric radii and velocity components: U (locally toward the center), V (in the direction of Galactic rotation), and W (toward the north Galactic pole).

\begin{table*}[htb]
\centering
\resizebox{\columnwidth}{!}{%
\begin{tabular}{lrrrrrrrrr} 
\toprule \toprule
\multicolumn{1}{c}{\bf Source} &
\multicolumn{1}{c}{ $\bf \pi$} &
\multicolumn{1}{c}{\bf $ \bf \mu_{x}$} &
\multicolumn{1}{c}{\bf $ \bf \mu_{y}$} &
\multicolumn{1}{c}{\bf R} &
\multicolumn{1}{c}{\bf V$ \bf _{LSR}$} &
\multicolumn{1}{c}{\bf U$_{s}$} &
\multicolumn{1}{c}{\bf V$_{s}$} &
\multicolumn{1}{c}{\bf W$_{s}$} &
\multicolumn{1}{c}{\bf Ref} \\
\multicolumn{1}{c}{\bf  Name} &
\multicolumn{1}{c}{\bf (mas)} &
\multicolumn{1}{c}{\bf (mas~y$^{-1}$)} &
\multicolumn{1}{c}{\bf (mas~y$^{-1}$)} &
\multicolumn{1}{c}{\bf (kpc)}   &
\multicolumn{1}{c}{\bf  (km~s$^{-1}$)}   &
\multicolumn{1}{c}{\bf  (km~s$^{-1}$)}   &
\multicolumn{1}{c}{\bf  (km~s$^{-1}$)}   &
\multicolumn{1}{c}{\bf  (km~s$^{-1}$)}   & \\
\hline\hline
G008.83$-$00.02    & $0.208\pm0.019$ & $-1.35\pm0.27$ & $-1.43\pm0.46$ & 3.48 & $ 01\pm05 $ & $ -39\pm11  $ & $197\pm09$  & $ 18\pm05$ & $^{a}$ \\
G009.62$+$00.19    & $0.194\pm0.023$ & $-0.58\pm0.13$ & $-2.49\pm0.29$ & 3.19 & $ 02\pm03 $ & $ -44\pm17  $ & $179\pm17$  & $-10\pm05$ & (1) \\        
G010.47$+$00.02    & $0.115\pm0.008$ & $-3.69\pm0.08$ & $-6.34\pm0.07$ & 1.63 & $ 68\pm05 $ & $ 04\pm31   $ & $124\pm24$  & $ 09\pm07$ & $^{a}$, (2) \\
G010.62$-$00.38    & $0.202\pm0.019$ & $-0.37\pm0.50$ & $-0.60\pm0.25$ & 3.41 & $-03\pm05 $ & $ -66\pm16  $ & $217\pm10$  & $ 08\pm11$ & (2) \\ 
G012.02$-$00.03    & $0.106\pm0.008$ & $-4.21\pm0.11$ & $-7.96\pm0.28$ & 2.24 & $109\pm05 $ & $  24\pm31  $ & $226\pm29$  & $ 01\pm07$ & (2) \\ 
G020.77$-$00.05    & $0.124\pm0.013$ & $-3.27\pm0.26$ & $-6.55\pm0.27$ & 2.92 & $ 57\pm03 $ & $  29\pm16  $ & $149\pm16$  & $ 02\pm10$ & (3) \\ 
G023.00$-$00.41    & $0.205\pm0.015$ & $-1.72\pm0.22$ & $-4.12\pm0.37$ & 4.13 & $ 79\pm05 $ & $  14\pm11  $ & $207\pm07 $ & $-01\pm06$ & (4,5) \\   
G023.38$+$00.18    & $0.208\pm0.025$ & $-1.57\pm0.23$ & $-3.92\pm0.26$ & 4.20 & $ 75\pm03 $ & $  06\pm17  $ & $208\pm09 $ & $-02\pm06$ & (1) \\ 
G023.43$-$00.18    & $0.170\pm0.032$ & $-1.93\pm0.15$ & $-4.11\pm0.13$ & 3.61 & $ 97\pm03 $ & $ -12\pm40  $ & $207\pm17$  & $ 02\pm04$ & (5) \\
G023.70$-$00.19    & $0.161\pm0.024$ & $-3.24\pm0.18$ & $-6.28\pm0.21$ & 3.51 & $ 76\pm05 $ & $  44\pm16  $ & $167\pm09$  & $ 07\pm06$ & (2) \\
G024.78$+$00.08    & $0.188\pm0.007$ & $-2.60\pm0.12$ & $-4.55\pm0.30$ & 4.00 & $110\pm05 $ & $  33\pm06  $ & $223\pm05$  & $ 14\pm01$ & $^{a}$ \\ 
G024.85$+$00.08    & $0.176\pm0.016$ & $-2.42\pm0.19$ & $-5.23\pm0.26$ & 3.83 & $111\pm05 $ & $  34\pm16  $ & $215\pm06$  & $ 01\pm06$ & (6) \\     
G026.50$+$00.28    & $0.159\pm0.012$ & $-2.51\pm0.34$ & $-6.04\pm0.34$ & 3.77 & $104\pm03 $ & $  33\pm15  $ & $206\pm04$  & $-08\pm10$ & (3) \\  
G028.14$-$00.00    & $0.158\pm0.023$ & $-2.11\pm0.17$ & $-4.85\pm0.17$ & 3.94 & $100\pm05 $ & $ -12\pm27  $ & $215\pm09$  & $-03\pm06$ & (7) \\  

\bottomrule
\bottomrule
\end{tabular}%
}
\caption{Detailed information on parallax and proper motion measurements of 14 massive and young stars with maser emission. Column (1): the Galactic name. Columns (2)--(4): the source parallax and proper motion in the eastward ($\mu_x$ = $\mu_\alpha\cos\delta$) and northward ($\mu_y$ =$\mu_\delta$) directions. Columns (5) and (6): the Galactocentric radius and LSR velocity of the star associated with the maser source. Columns (7)--(9): the Galactocentric velocities of the source towards the Galactic center (U$_{s}$), in the direction of the Galactic rotation (V$_{s}$) and perpendicular to the Galactic plane (W$_{s}$) respectively. Column (10): publication references: (1)\cite{Sanna:2009} ; (2)\cite{Sanna:2014}; (3)\cite{Xu:2021}; (4)\cite{Reid:2019}; (5)\cite{Brunthaler:2009}; (6)\cite{JLi:2022}; (7)\cite{Immer:2019}. $^{a}$ This work.
}
\label{tab:14BarSources}
\end{table*}

Note that these (U, V, W) velocities are not peculiar motions, i.e., they do not have a model of Galactic rotation subtracted. Instead, they are heliocentric motions, with the full orbital motion of the Sun added back, in order to place them in a Galactocentric frame of reference. The fundamental parameters used to calculate Galactocentric motions are as follows: distance to the Galactic center(GC) $R_{o}$ = 8.15~kpc; the circular rotation speed at the Sun's position $\Theta_{o}$ = 236~km~s$^{-1}$; the solar motion toward GC $U_{\odot}$ = 10.6 km~s$^{-1}$; the solar motion in direction of Galactic rotation $V_{\odot}$ = 10.7 km~s$^{-1}$; and the solar motion towards north Galactic pole $W_{\odot}$ = 7.6 km~s$^{-1}$ \citep{Reid:2019}.

\clearpage

\section{Results and Discussion}\label{sec:ResultsDiscussion}

\begin{figure}[H]
\centering
    \begin{subfigure}
        \centering
        \begin{overpic}[width=0.6\linewidth]{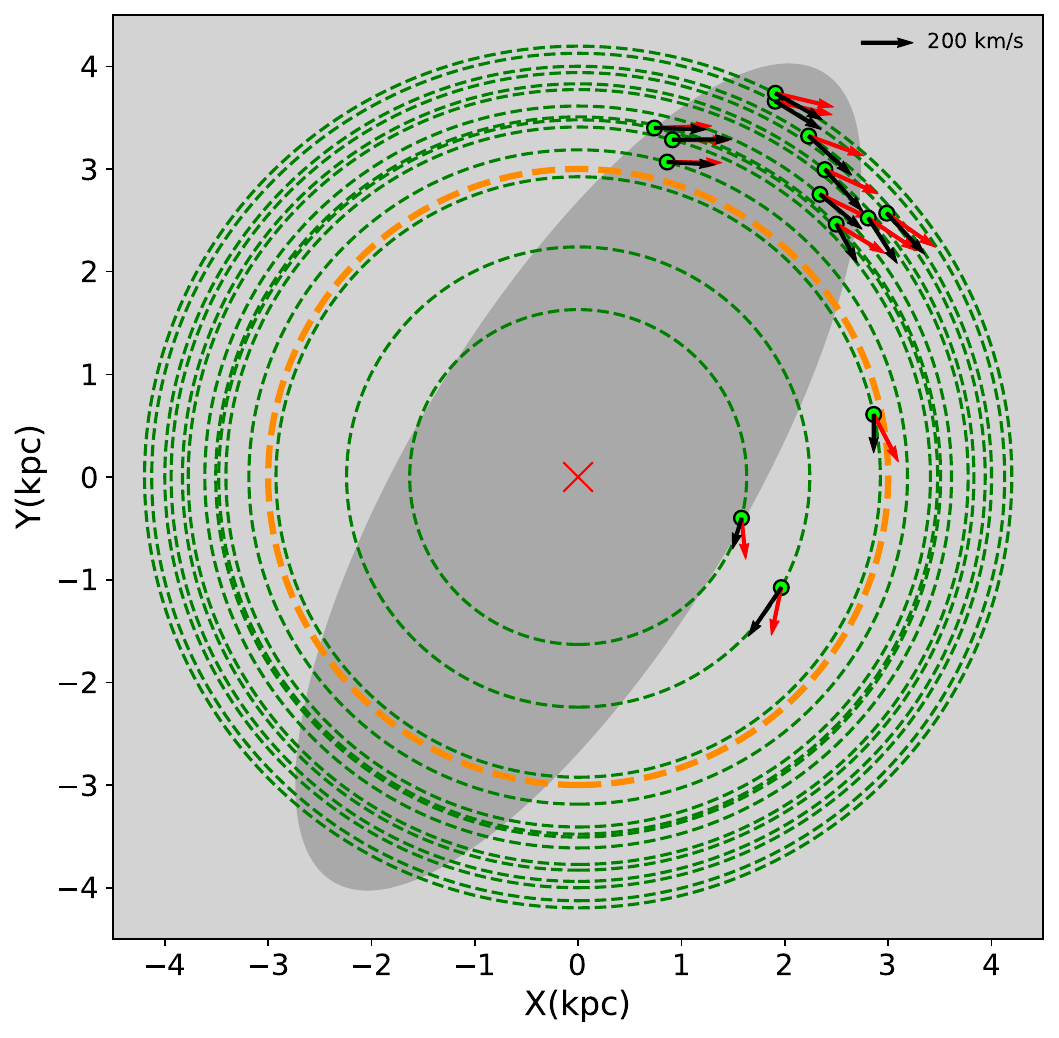}
            \put(12,92){\large (a)}  
        \end{overpic}
        \label{fig:ExpandingArmsModel_PV}
    \end{subfigure}
    \hfill
    \begin{subfigure}
        \centering
        \begin{overpic}[width=0.72\linewidth]{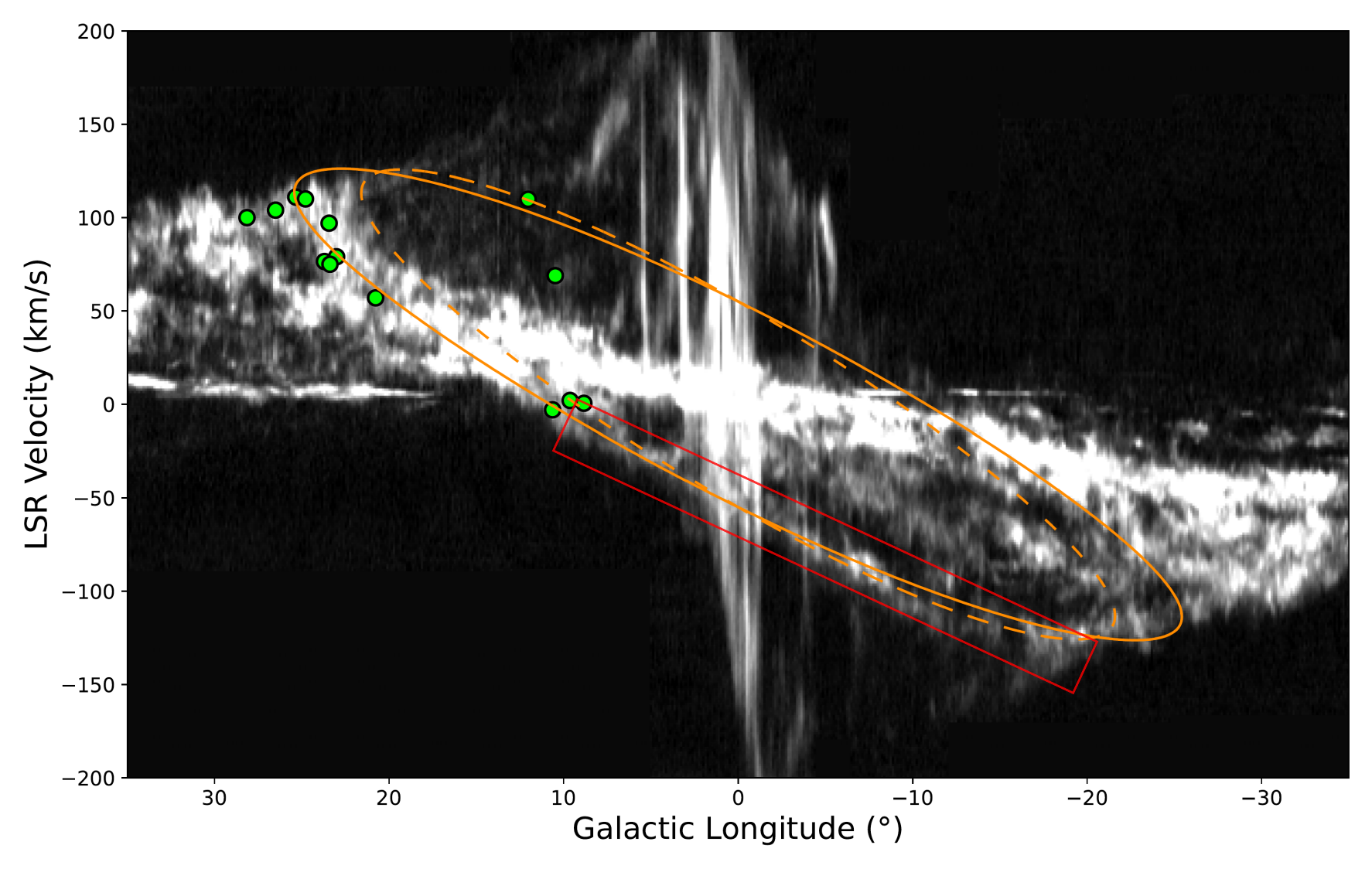}
            \put(10,58){\textcolor{white}{\large (b)}}
        \end{overpic}
        \label{fig:ExpandingArmsModel_LV}
    \end{subfigure}
\caption{Positions and motions of massive young stars (green dots) in the inner Milky Way, based on VLBA parallax and proper motion measurements of associated masers. (a) A plan view from the north Galactic pole.  The red `$\times$' locates the Galactic center, with the Sun off the top of the figure at (X,Y)=(0,8.2) kpc. The gray ellipse represents the Galactic bar \citep{Wegg:2015}. The dashed orange circle has a radius of 3 kpc. The dashed green circles have radii set to match a star's distance from the Galactic center. Black arrows indicate our measured Galactocentric velocities. Red arrows show the Galactocentric velocities of these stars predicted by the Expanding 3 kpc arms model, assuming values for the fundamental Galactic parameters of $R_{0}$ = 8.15~kpc and $\Theta_{0}$ = 236~km s$^{-1}$ and an expansion velocity of 55~km s$^{-1}$. (b) An $l-v$ plot of CO emission in the Milky Way \citep{Dame:2001} integrated over Galactic latitude of $\pm$1$^\circ$. The dashed orange line traces the Expanding 3 kpc arms; the solid orange line represents an alternative model for the Expanding 3 kpc arm, adopting R = 3.5 kpc and a circular rotation of 215~km s$^{-1}$. The red rectangle marks prominent CO emission associated with the Near 3 kpc arm. The green dot at (25.35,111) has been shifted by $+$0.5$^\circ$ in longitude from its original position for clarity.
}
\label{fig:ExpandingArmsModel}
\end{figure}

We analyze distances and three-dimensional (3D) motions for 14 newly formed massive stars within 4.2 kpc of the Galactic center. Figure~\ref{fig:ExpandingArmsModel} shows their locations projected onto the Galactic plane (Figure~\ref{fig:ExpandingArmsModel}(a)) and in an $l - v$ diagram of carbon monoxide (CO) emission, which traces interstellar molecular gas (Figure~\ref{fig:ExpandingArmsModel}(b)). Their motions in a reference frame at rest at the Galactic center are indicated by black arrows. The motion of the sources results in a discernible pattern resembling elliptical orbits around the Galactic center.

A dashed orange circle in Figure~\ref{fig:ExpandingArmsModel}(a) represents a historical Expanding 3 kpc arms model \citep{Bania:1977}, assuming an expansion velocity of 55~km s$^{-1}$ superposed on a circular motion of 200 km s$^{-1}$ and 3~kpc radius. The model uses the same fundamental Galactic parameter values as given above. We adopted the orbital speed of 200 km s$^{-1}$ as a round number that is close to the value determined from the rotation curve of the Milky Way at 3 kpc of 203 km s$^{-1}$ from \cite{Reid:2019}. 

Figure~\ref{fig:ExpandingArmsModel}(a) shows that the historical Expanding 3 kpc arms model (dashed orange circle) fails to fit most of our sources in both positions and motions. This suggests that the arms might not be at 3~kpc radius and hints at the possibility that the sources may be moving in noncircular orbits.

To enable a quantitative comparison with our measurements, we consider a model where the Expanding 3 kpc arms are made up of multiple circular orbits with individual radii. Such a scenario is shown in Figure~\ref{fig:ExpandingArmsModel}(a) by dashed green circles. The motions of the stars predicted by the expanding arms model are shown by red arrows. Analysis of Figure~\ref{fig:ExpandingArmsModel} reveals that this model provides a poor match to the velocity vectors (both in magnitude and direction). Our sources tend to move inwards (towards the Galactic center) faster than the model predicts. We compared the observed motions in the x and y directions with the corresponding motions predicted by the model. The uncertainties in the x and y motions were derived through error propagation based on the uncertainties in U and V in Table~\ref{tab:14BarSources}. This comparison yielded a reduced chi-squared ($\chi^{2}_{\nu}$) per degree of freedom equal to 43.3, with the analysis involving 14 degrees of freedom.

The $l - v$ diagram in Figure~\ref{fig:ExpandingArmsModel}(b) shows a historical model for the Expanding 3 kpc arms \citep{Bania:1977} with a single 3~kpc radius (dashed orange line). This model covers the CO emission regions between 10$^\circ$ to $-$20$^\circ$ Galactic longitudes (highlighted by the red rectangle), which have been associated with the Near 3 kpc arm since its discovery \citep{vanWoerden:1957}. The model does a good job of matching three of our maser sources near Galactic longitude $10^\circ$, which are located in the CO emission that is historically been associated with the Near 3 kpc arm. The maser sources beyond Galactic longitude $20^\circ$ tend to lie outside the Expanding 3 kpc arms model. However, since spiral arms have significant widths, we also plot an expanding arms model with a radius of 3.5 kpc (solid-orange), which then includes those maser sources.

The orbits of stars in the gravitational potential of a bar have been theoretically analyzed and simulated (e.g., \cite{ContopoulosPapayannopoulos:1980}). In brief, there are two families of concentric resonant orbits about a bar: ``X1'' orbits with semi-major axes parallel to the bar, and ``X2'' orbits perpendicular to the bar. The superposition of these orbits along with other gas orbits trapped inside them form what is termed the bar. The positions and motions of our stars are a reasonable match to a range of X1 orbits around the Galactic center.

\begin{figure}[H]
    \centering
    \begin{subfigure}
        \centering
        \begin{overpic}[width=0.6\linewidth]{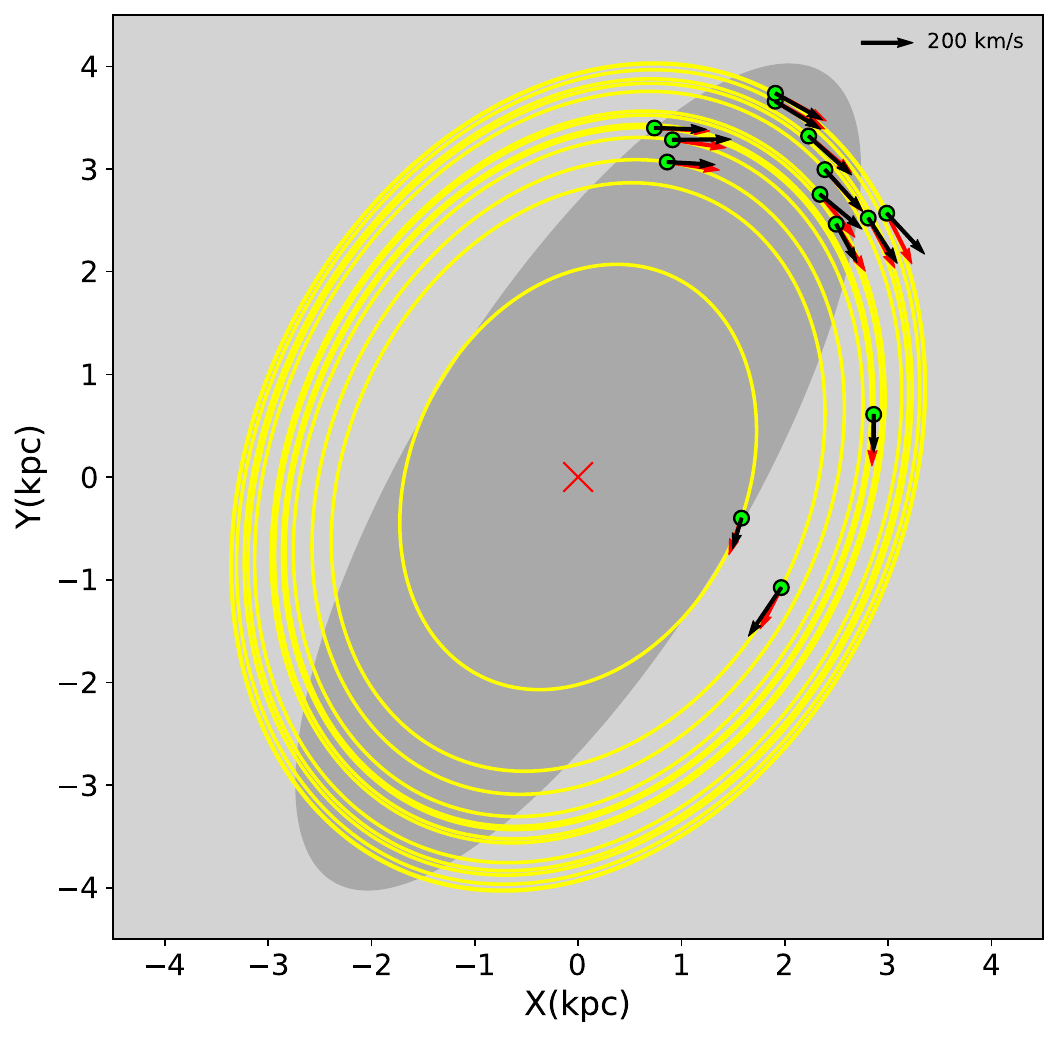}
            \put(12,92){\textcolor{black}{\large (a)}}
        \end{overpic}
        \label{fig:X1OrbitsModel_PV}
    \end{subfigure}
    \hfill
    \begin{subfigure}
        \centering
        \begin{overpic}[width=0.8\linewidth]{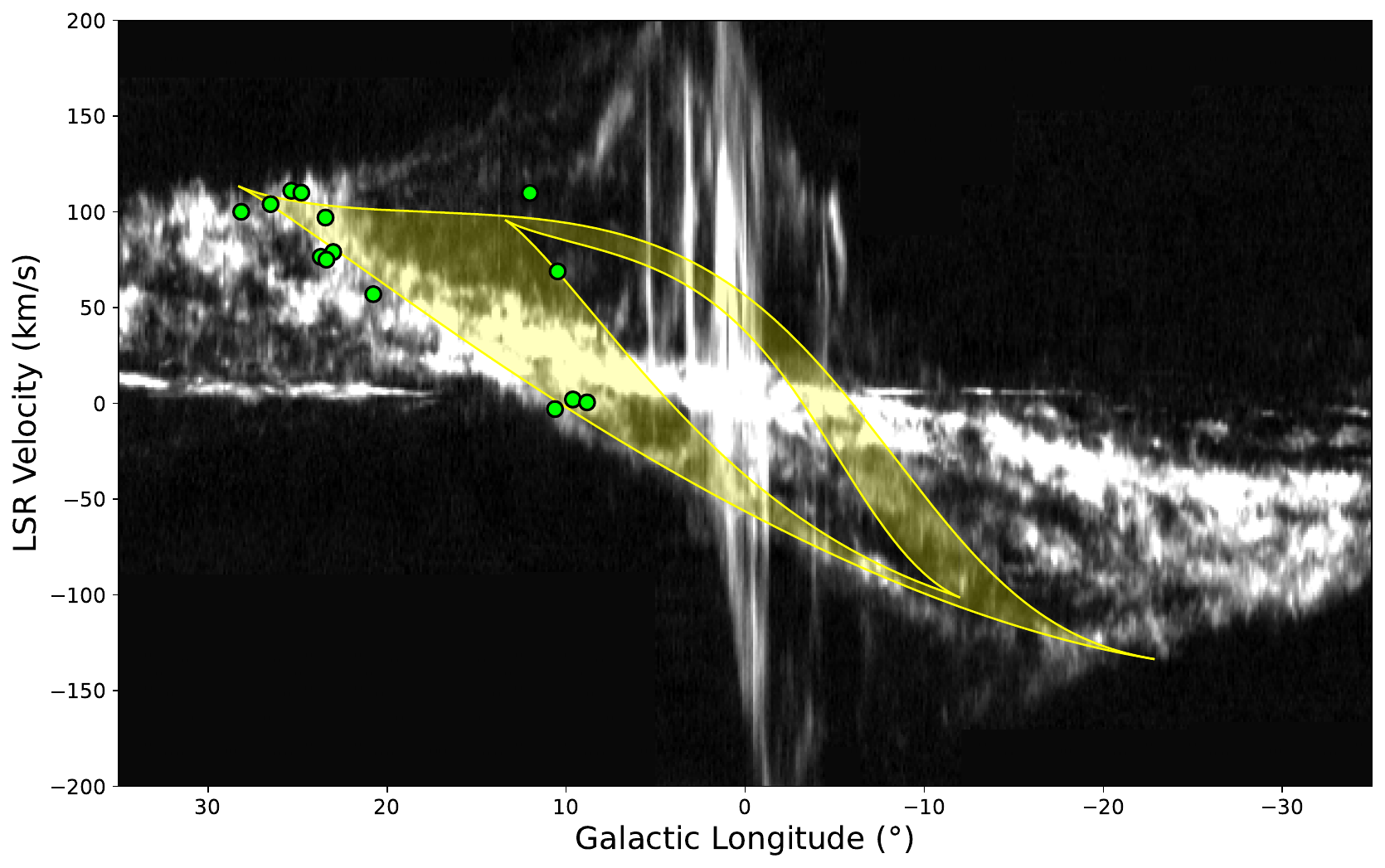}
            \put(10,58){\textcolor{white}{\large (b)}}
        \end{overpic}
        \label{fig:X1OrbitsModel_LV}
    \end{subfigure}

\caption{Quasi-elliptical (X1) orbits about the Galactic bar. (a) Our best-fit models showing a range of concentric elliptical X1 orbits about the Galactic bar, which provide a good match to the motions of massive young stars (see Figure~\ref{fig:ExpandingArmsModel} caption for details). (b) The range of these models in an $l-v$ diagram (solid yellow). Note that our models fit the parallax and proper motion data well and follow the quasi-linear lane of CO emission between ($l,v$) = (10$^{\circ}$,0 km s$^{-1}$) and ($-20^{\circ}$,$-120$ km s$^{-1}$) previously associated with the Near 3 kpc arm.
}
\label{fig:X1OrbitsModel}
\end{figure}

We propose that gas and stars between 2.2 and 4.2 kpc of the Galactic center orbit about the bar, and the structures previously identified as the Expanding 3 kpc arms are segments of these orbits. Our model provides for a range of concentric elliptical orbits, approximating a family of X1 orbits, around the Galactic bar, represented by solid yellow lines in Figure~\ref{fig:X1OrbitsModel}(a). 

Our proposed model can qualitatively explain the apparent expansion in velocities seen along the line of sight through the Galactic center. In Figure~\ref{fig:X1OrbitsModel}(a), for clockwise motions along the near side of the ellipse, a star (or gas) would have a component of its orbital motion directed toward the Sun, whereas material on the far side would have a component of motion directed away from the Sun, as observed. 

In order to make a quantitative comparison of our model with the observations, we performed a Markov Chain Monte Carlo (MCMC) fitting, where we varied the Galactocentric azimuth (measured clockwise from the Sun as viewed from the north Galactic pole) and semimajor to semiminor axis ratio for the concentric ellipses to minimize the difference between the measured motions and motions predicted by our model. For each MCMC trial, we set the semimajor axis of the ellipse to make it pass through the point. 
We used the rotation curve of the Milky Way from \cite{Reid:2019} to provide an approximation of gravitational potential and set the full motion along an ellipse as the circular rotation velocity at the radius $R$ to the maser. For example, for a given maser point at (1,-2), a trial Galactocentric azimuth of 45$^{\circ}$ and axis ratio of 1.3 require an X1 orbit with a semimajor axis of 1.84, which runs through the maser point. We then calculate the velocity along the ellipse. The 2D marginal and joint parameter distributions from the MCMC fitting are shown in Figure~\ref{fig:appA10} in Appendix~\ref{sec:appendixA}. Details about the X1 orbit modeling are provided in  Appendix~\ref{sec:appendixC}.

Our best-fit elliptical X1-orbits model gives the Galactocentric azimuth of the bar to be $25^{+6^\circ}_{-9^\circ}$, with a semimajor to semiminor axis ratio of $1.33^{+0.09}_{-0.19}$. The $\chi^{2}_{\nu}$\ per degree of freedom calculated for our best-fit model is 4.5, which is calculated with the same number of degrees of freedom (14) as employed in the expanding arms model. This represents a significant improvement over the $\chi^{2}_{\nu}$\ value calculated for the expanding arms model discussed earlier. The extent and orientation of the elliptical annulus are in reasonable agreement with the parameters of the Galactic bar estimated from multiple studies\footnote{In the literature there is discussion of a short bar and a long bar \citep{Hammersley:2000, Benjamin:2005, Athanassoula:2006}, but since our stars closely match the long bar's dimensions and orientation, we assume the two are closely related.}, e.g., near-infrared star counts, red clump giant measurements around the Galactic center, and the Galactic Legacy Mid-Plane Survey Extraordinaire \citep{Hammersley:1994, Hammersley:2000, Benjamin:2005, Cabrera-Lavers:2007, Lopez-Corredoira:2007, Cabrera-Lavers:2008, Zasowski:2012, Wegg:2015}.

Figure~\ref{fig:X1OrbitsModel}(b) shows the same $l - v$ diagram of CO emission as in Figure~\ref{fig:ExpandingArmsModel}(b). But instead of the historical model for the Expanding 3 kpc arms, here we present a hatched elliptical annulus (shown in solid yellow) based on our best-fit model of elliptical X1 orbits. One can see that the measured locations of almost all of our stars lie very close to the model.

Figure~\ref{fig:ExpandingArmsModel}(b) and ~\ref{fig:X1OrbitsModel}(b) also present a more detailed comparison of the motions expected for the Expanding 3 kpc arms model and elliptical bar orbits. As shown in Figure~\ref{fig:X1OrbitsModel}(b), our proposed model demonstrates a relatively good match with the CO emission that was previously attributed to the near side of the Expanding 3 kpc arms. It does a good job of matching the far side of the Expanding 3 kpc arm, although that arm is hard to see in these figures (see Figure 1 of \cite{Dame:2008} and Figure 14 of \cite{Reid:2016}). 

By its very nature, the Expanding 3 kpc arms model requires elliptical orbits, since a radial velocity component is involved. One can also see that an elliptical annulus shown in Figure~\ref{fig:X1OrbitsModel}(b) can mimic the signatures of expansion that led to the early historical Expanding arms model. This supports our proposal that the Expanding 3 kpc arms are, in fact, parts of bound orbits about the bar and not expanding from the Galactic center. 

\begin{figure}[H]
\centering
\includegraphics[width=0.95\textwidth]{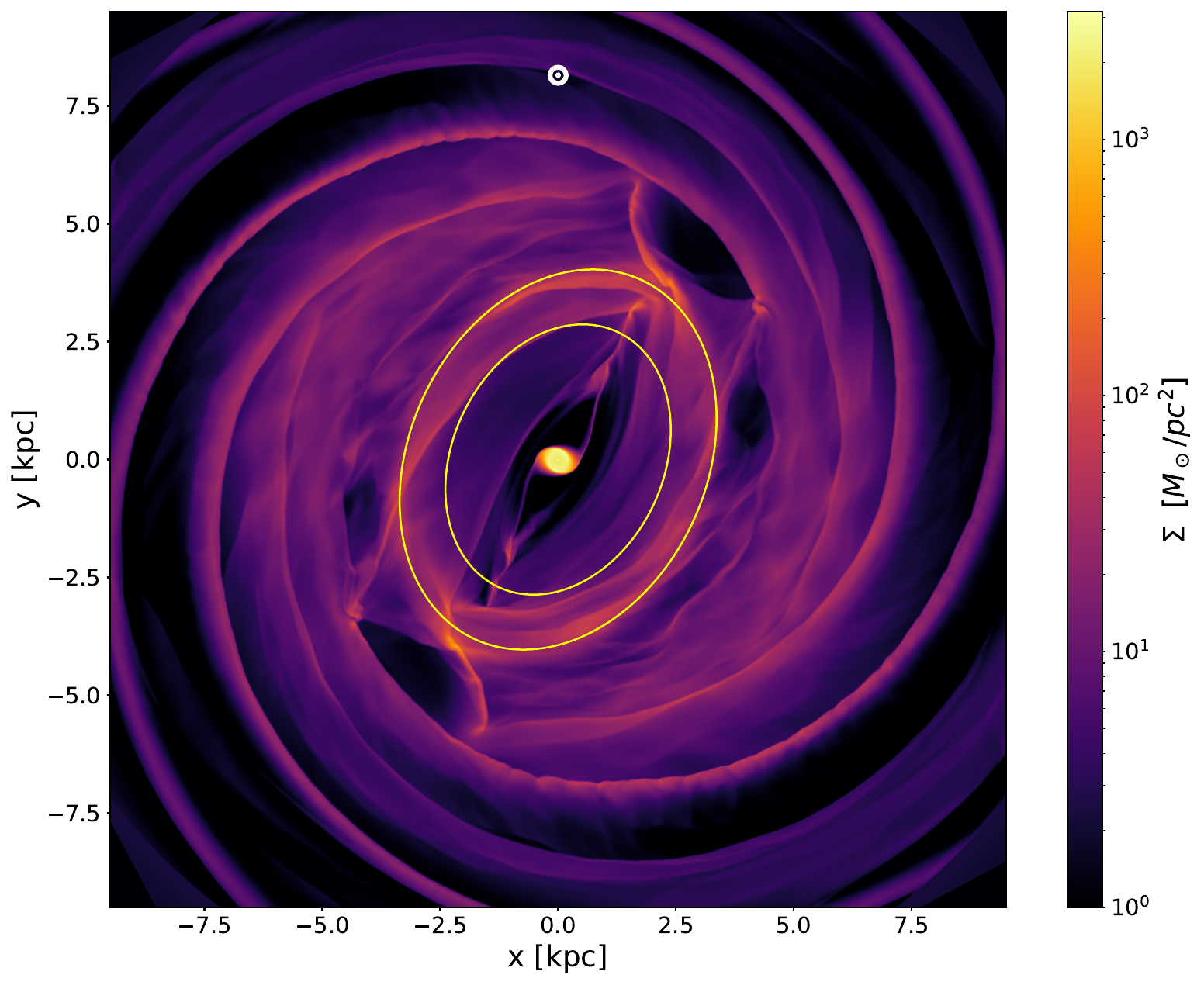}
\caption{Our proposed model showing the range of X1 orbits about the Galactic bar (represented by the area enclosed by yellow ellipses that trace the orbits of stars about the Galactic bar) overplotted on a simulation of the gas surface-density of the Milky Way \citep{Li:2022}. The Sun (white circle) is at the top. The color bar on the right indicates surface density. This shows a qualitative agreement between the simulated gas density and an elliptical orbit about the Galactic bar.
}
\label{GasSDLiModel}
\end{figure}

The masers we have measured are associated with massive stars that are very young ($\ll1$ Myr old) and hence closely track the gas clouds out of which they formed. A hydrodynamic simulation of the distribution of gas in the Milky Way can be compared to our results. The simulation in Figure~\ref{GasSDLiModel}, taken from \cite{Li:2022}, displays gas surface density and shows that high-density gas near the Galactic bar can form a range of quasi-elliptical structures similar to the elliptical annulus shown in Figure~\ref{fig:X1OrbitsModel}(a). 

The maser with the smallest semimajor axis (2.2 kpc) might be in an X2 orbit perpendicular to the bar. If we exclude this orbit, the annular region predicted by our model for X1 orbits is outlined by the solid yellow ellipses in Figure~\ref{GasSDLiModel}. This is in good agreement with the shapes of the density peaks around the Galactic center in the simulation. Note that there is a thin, nearly elliptical ridge just on the inner boundary of the proposed model range and, interestingly, a quasi-linear structure with a bend in it just inside the other boundary (at Galactic longitudes around $5^{\circ}-15^{\circ}$). This quasi-linear ridge structure would have near-zero LSR velocity values (since the ridge line runs nearly sideways), supporting what we found for sources on the near side of the bar orbits (sources near (1,3)~kpc in Figure~\ref{fig:X1OrbitsModel}(a)). This agreement supports our suggestion that the Expanding 3 kpc arms are actually segments of elliptical orbits about the Galactic bar. We plan future VLBI observations to locate and determine the motions of massive young stars with masers near the far end of the bar to critically test our model and improve the parameters of the elliptical orbits.

Finally, we note that our new understanding of the 3 kpc arms leads to a revised model for the Norma spiral arm. The most recent summary paper of the BeSSeL Survey \citep{Reid:2019} introduced a kink in their model of the Norma arm near where it approaches the end of the Galactic bar (indicated by a 'K' in their Figure 1). The kink allowed for some massive young stars with masers to bend inward (clockwise) from the end of the bar. However, we find these stars to be located along the orbit about the bar and, thus, likely not associated with the Norma arm.  So in our proposed model, the Norma arm terminates near the end of the bar.

\section{Conclusions}\label{sec:conclusion}

The nature of the Expanding 3 kpc spiral arms has been debated since 1975, without a clear explanation for its apparent expansion. With accurate distance and proper motions from VLBA observations of masers associated with 14 massive young stars, we have shown that the apparent expansion instead can be well explained by a family of quasi-elliptical X1 orbits about the Galactic bar. We have provided quantitative evidence that these massive young stars are tracing these X1 orbits, rather than following the motions predicted by the Expanding 3 kpc arms model. In conclusion, the stars and gas which have been identified as forming the Expanding 3 kpc arms are neither expanding nor at a constant 3 kpc radius, nor spiral arms. They are in fact, as our title suggests, features of X1 orbits driven by the Galactic bar.

\section{Acknowledgments}\label{sec:acknowledge}
We thank Zhi Li, SHNU, Shanghai, China (\url{lizhi@shnu.edu.cn}) for providing the surface density model map in Figure~\ref{GasSDLiModel}. This project was developed in part at the Lorentz Center workshop Mapping the Milky Way, held 6-10 February 2023 in Leiden, the Netherlands. This work was supported by the Australian Research Council (ARC) Discovery grant Nos. DP180101061 and DP230100727.

\clearpage

\appendix

Here, we provide additional details regarding our observations and parallax measurements of the three H$_2$O maser sources discussed in Section~\ref{sec:2} and the MCMC fitting described in Section~\ref{sec:ResultsDiscussion}. Appendix~\ref{sec:appendixA} presents the interferometric spectrum, parallax fitting results of the maser sources, residual (internal) motion spot maps of the maser emission, and MCMC fitting results for the X1 orbit model; Appendix~\ref{sec:appendixB} includes four tables detailing the observational and calibration parameters, various parallax and proper motion fitting outcomes, and final results; and Appendix~\ref{sec:appendixC} gives details about the X1 orbit modeling.

To compute the residual motions shown in Figures~\ref{fig:appA7}-~\ref{fig:appA9}, at first, the proper motion of each maser feature was determined by averaging the motions of all associated maser spots. Then, the final values of proper motions in R.A. and decl. were obtained by averaging the motions of multiple maser features. Ultimately, these values were then subtracted from the motion of each individual maser spot to compute the residual motions. It is important to note that, in the absence of detailed knowledge about the location of the center of expansion, one could, in principle, add a single vector to all the motions.

\section{Figures}\label{sec:appendixA}

\renewcommand{\thefigure}{A\arabic{figure}} 
\setcounter{figure}{0} 

\begin{figure}[htp]
\centering
\includegraphics[angle=0,scale=0.5]{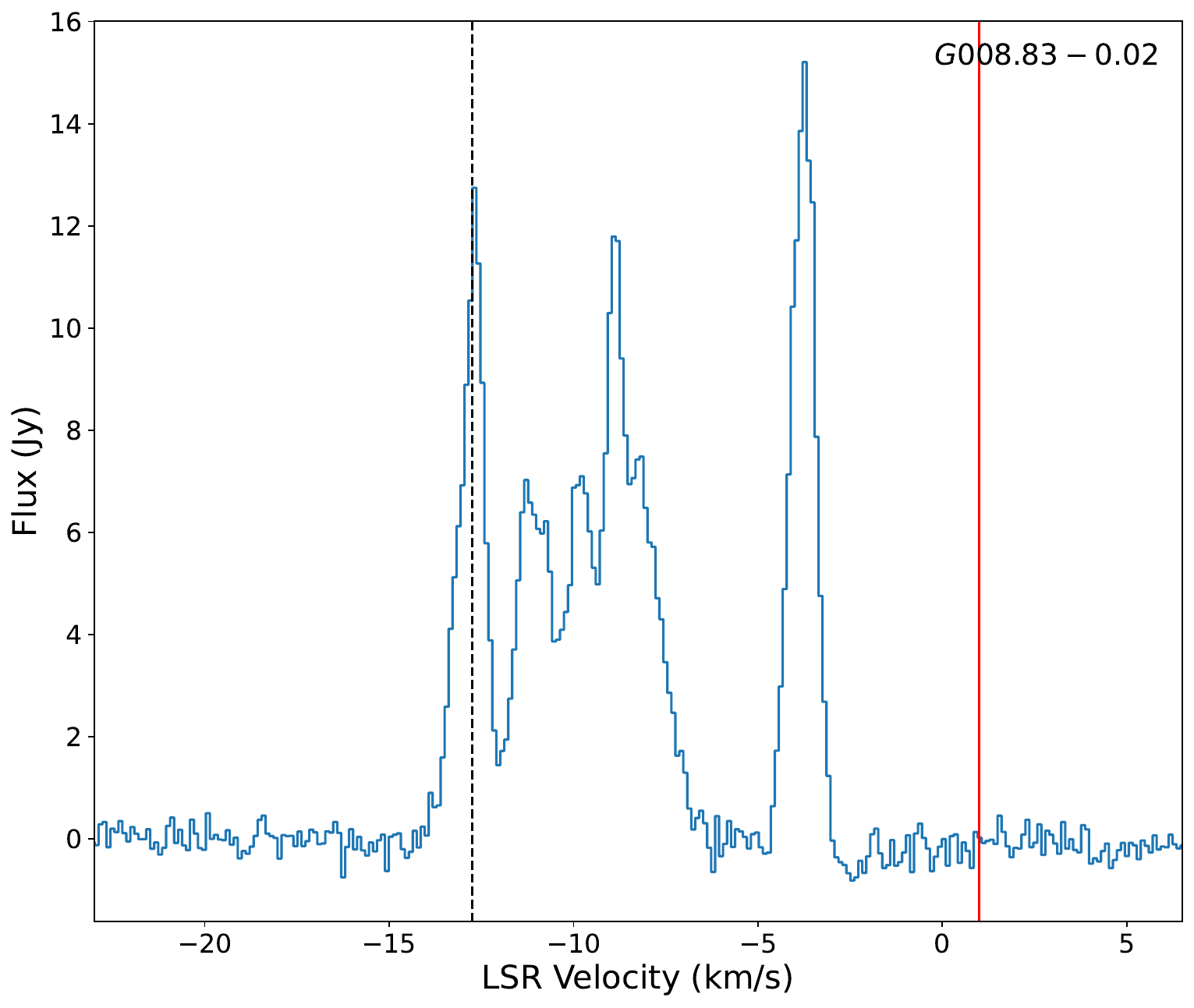}
\caption{Interferometric spectrum of the water maser emission for G008.83$-$00.02. The dashed vertical line indicates the LSR velocity of the maser spot used as the interferometric phase reference.  The red vertical line indicates the LSR velocity of the maser host star inferred from NH$_{3}$ $(J, K)$ = (1, 1) emission, which traces high-density molecular gas surrounding it \citep{Molinari:1996}.
}
\label{fig:appA1}
\end{figure}
\clearpage

\begin{figure}[htp]
\centering
\includegraphics[angle=0,scale=0.5]{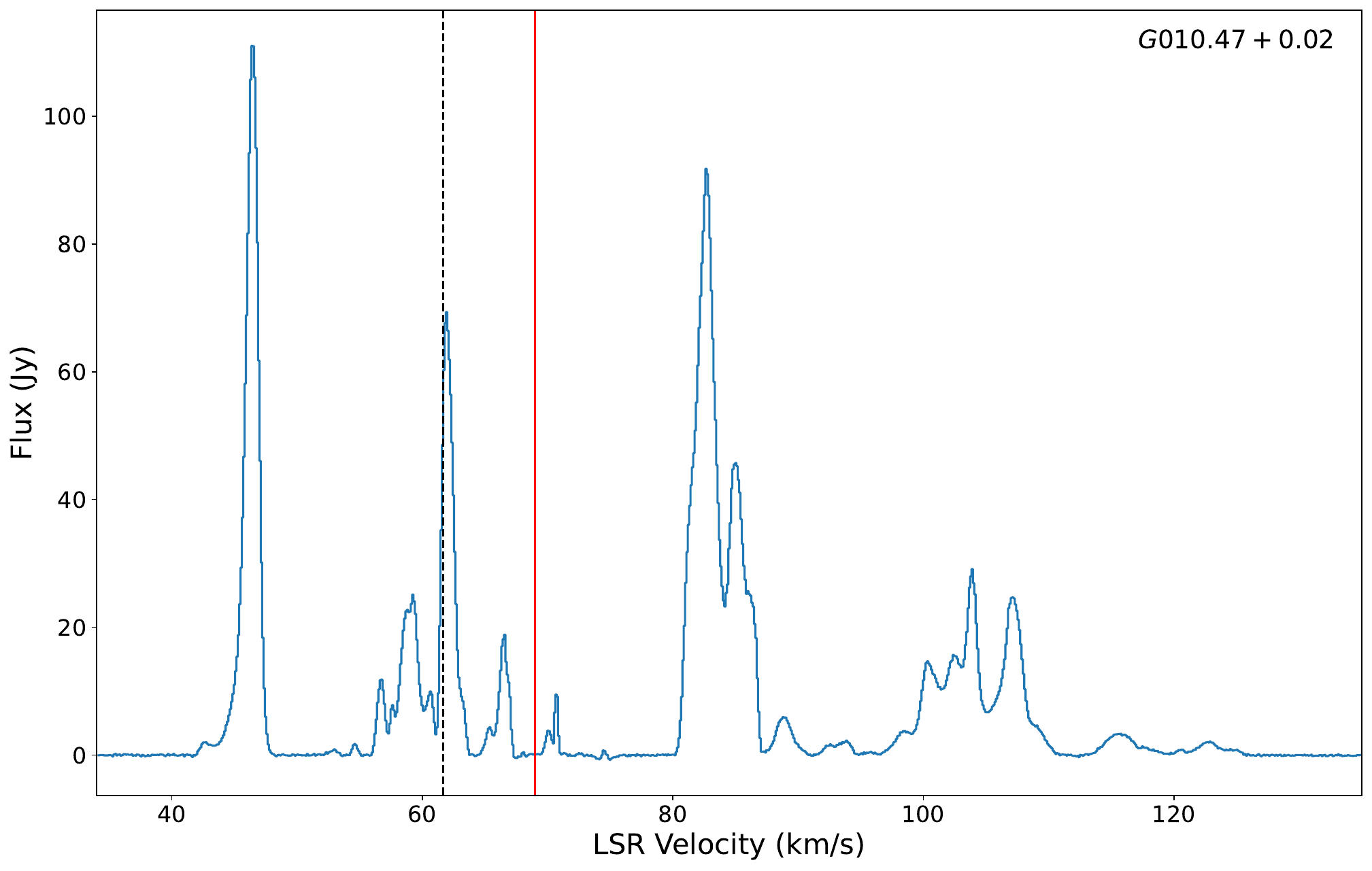}
\caption{Interferometric spectrum of the water maser emission for G010.47$+$00.02. The dashed vertical line indicates the LSR velocity of the maser spot used as the interferometric phase reference.  The red vertical line indicates the LSR velocity of the maser host star inferred from CS(2$-$1) emission, which traces high-density molecular gas surrounding it \citep{Bronfman:1996}.
}
\label{fig:appA2}
\end{figure}
\clearpage

\begin{figure}[htp]
\centering
\includegraphics[angle=0,scale=0.5]{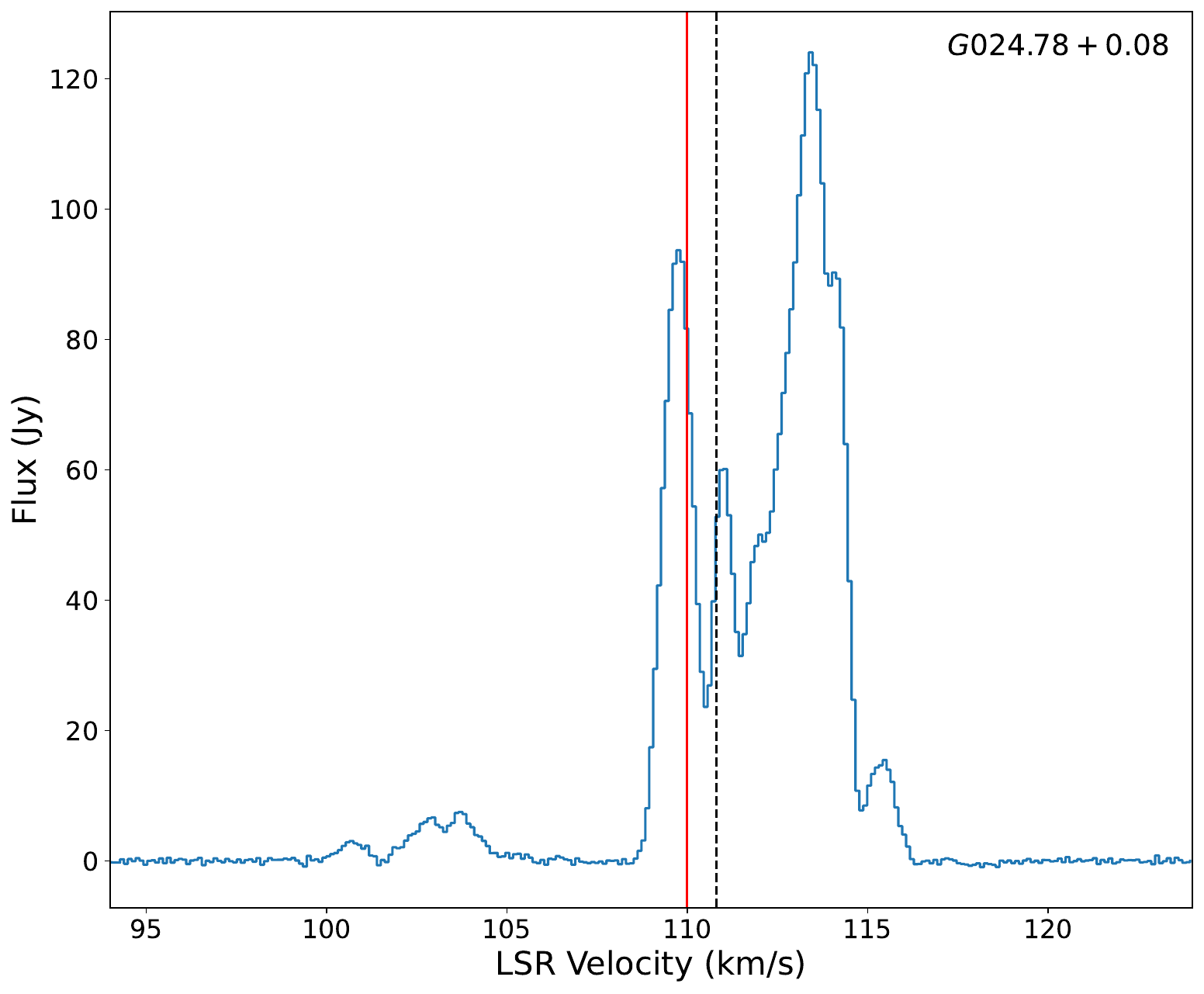}
\caption{Interferometric spectrum of the water maser emission for G024.78$+$00.08. The dashed vertical line indicates the LSR velocity of the maser spot used as the interferometric phase reference.  The red vertical line indicates the LSR velocity of the maser host star inferred from CS(1$-$0) emission, which traces high-density molecular gas surrounding it \citep{Anglada:1996}.
}
\label{fig:appA3}
\end{figure}
\clearpage

\begin{figure}[htp]
\centering
\includegraphics[angle= 0, scale= 0.60, trim=0 70 0 0, clip]{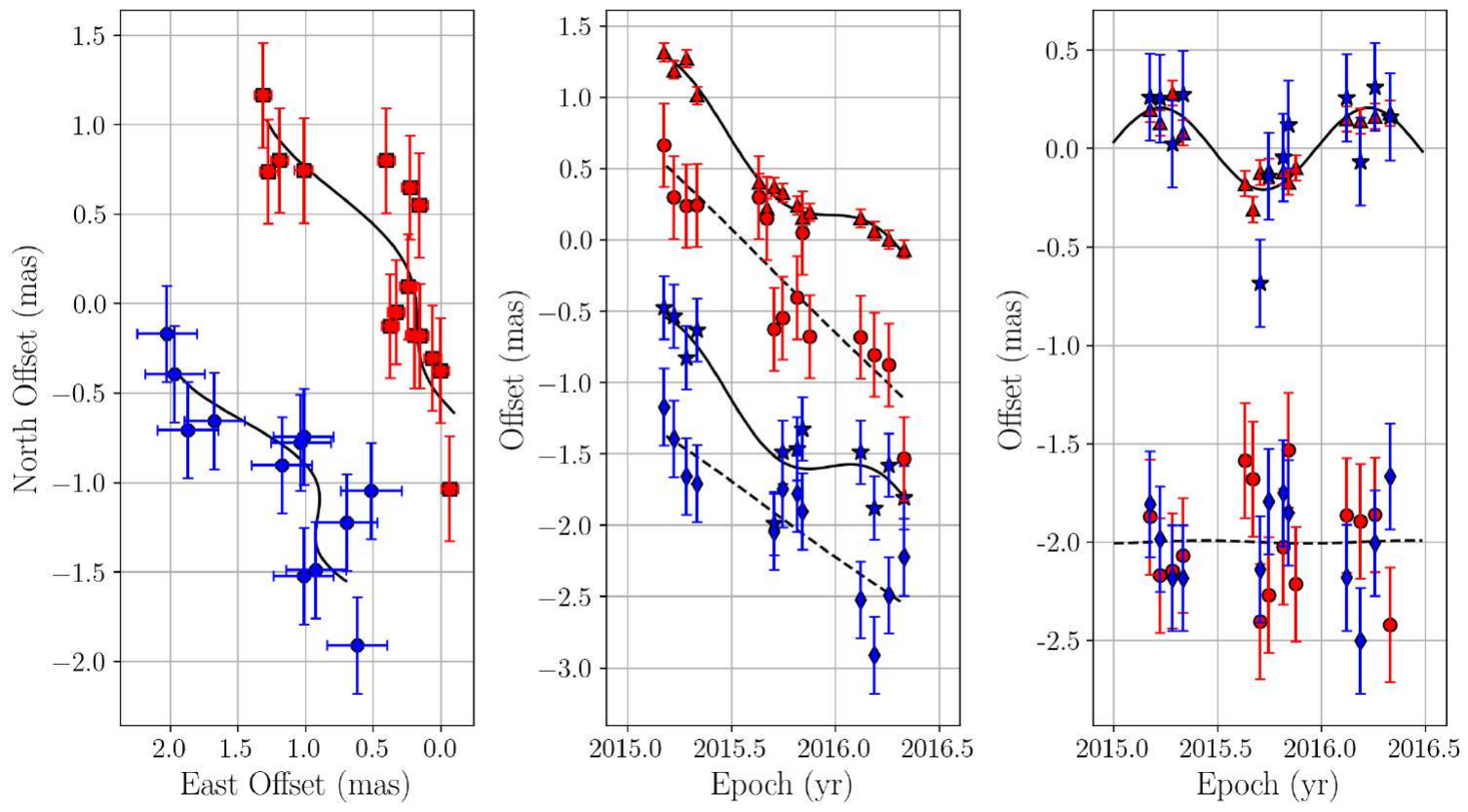}
\caption{Parallax and proper motion fits for the $-$12.74~km s$^{-1}$  maser spot of G008.83$-$00.02 with respect to calibrators J1755$-$2232 (red) \& J1813$-$2015 (blue).  Different position components and calibrators are offset for clarity. \textit{Left Panel:} position offsets on the sky with respect to the two calibrators are shown by squares and octagons. \textit{Middle Panel:} position offsets and best fits in the easterly (triangles and stars and solid lines) and northerly (dots and diamonds and dashed lines) directions vs. time. \textit{Right Panel:} same as the middle panel, but with the fitted proper motion subtracted (i.e., displaying only the parallax effect).
}
\label{fig:appA4}
\end{figure}
\clearpage

\begin{figure}[htp]
\centering
\includegraphics[angle= 0, scale= 0.60, trim=0 70 0 0, clip]{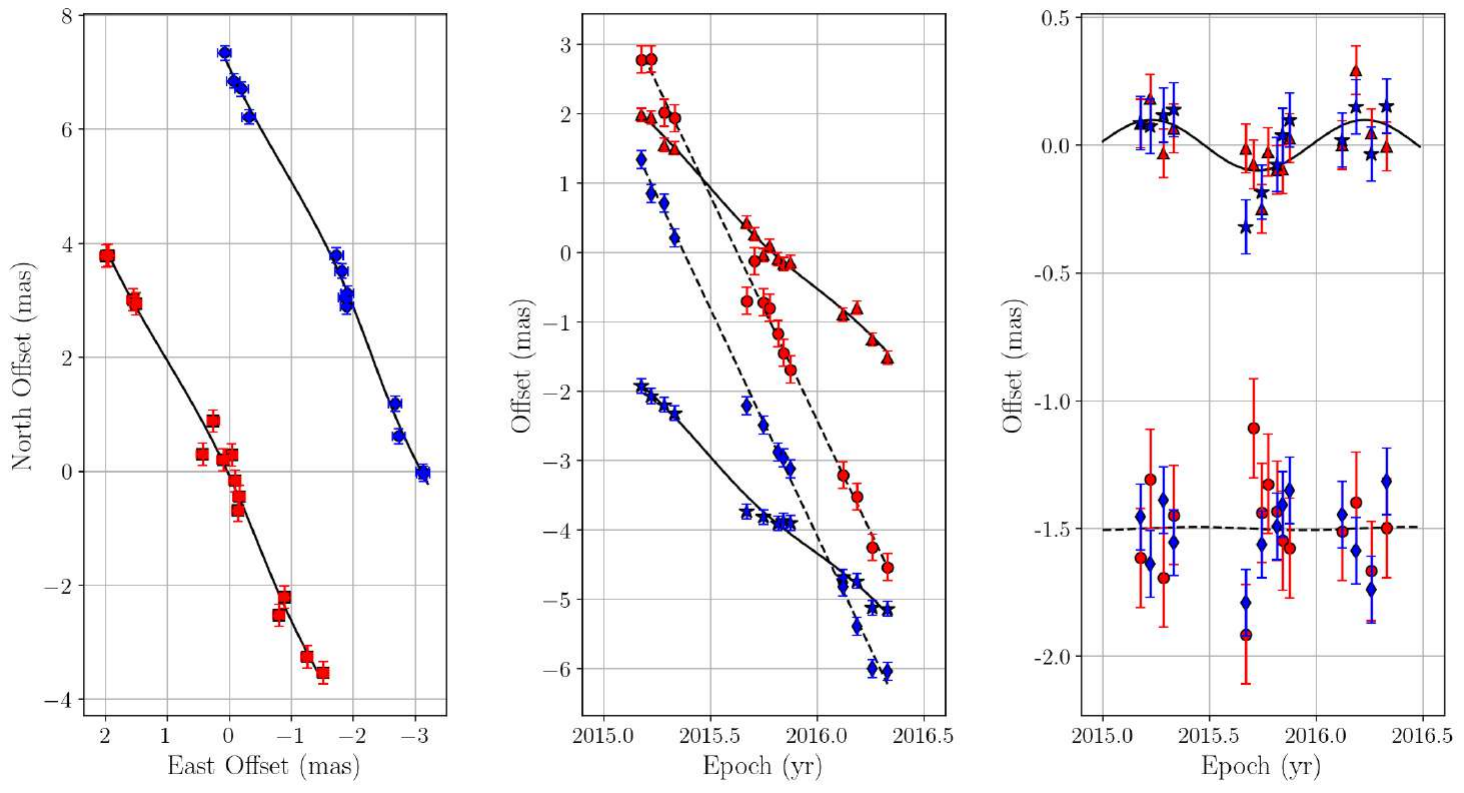}
\caption{Parallax and proper motion fits for the 61.68~km s$^{-1}$  maser spot of G010.47$+$00.02 with respect to calibrators J1808$-$1822 (red) \& J1813$-$2015 (blue). Different position components and calibrators are offset for clarity. \textit{Left Panel:} position offsets on the sky with respect to the two calibrators are shown by squares and octagons. \textit{Middle Panel:} position offsets and best fits in the easterly (triangles and stars and solid lines) and northerly (dots and diamonds and dashed lines) directions vs. time. \textit{Right Panel:} same as the middle panel, but with the fitted proper motion subtracted (i.e., displaying only the parallax effect).
}
\label{fig:appA5}
\end{figure}
\clearpage

\begin{figure}[htp]
\centering
\includegraphics[angle=0, scale=0.60, trim=0 50 0 0, clip]{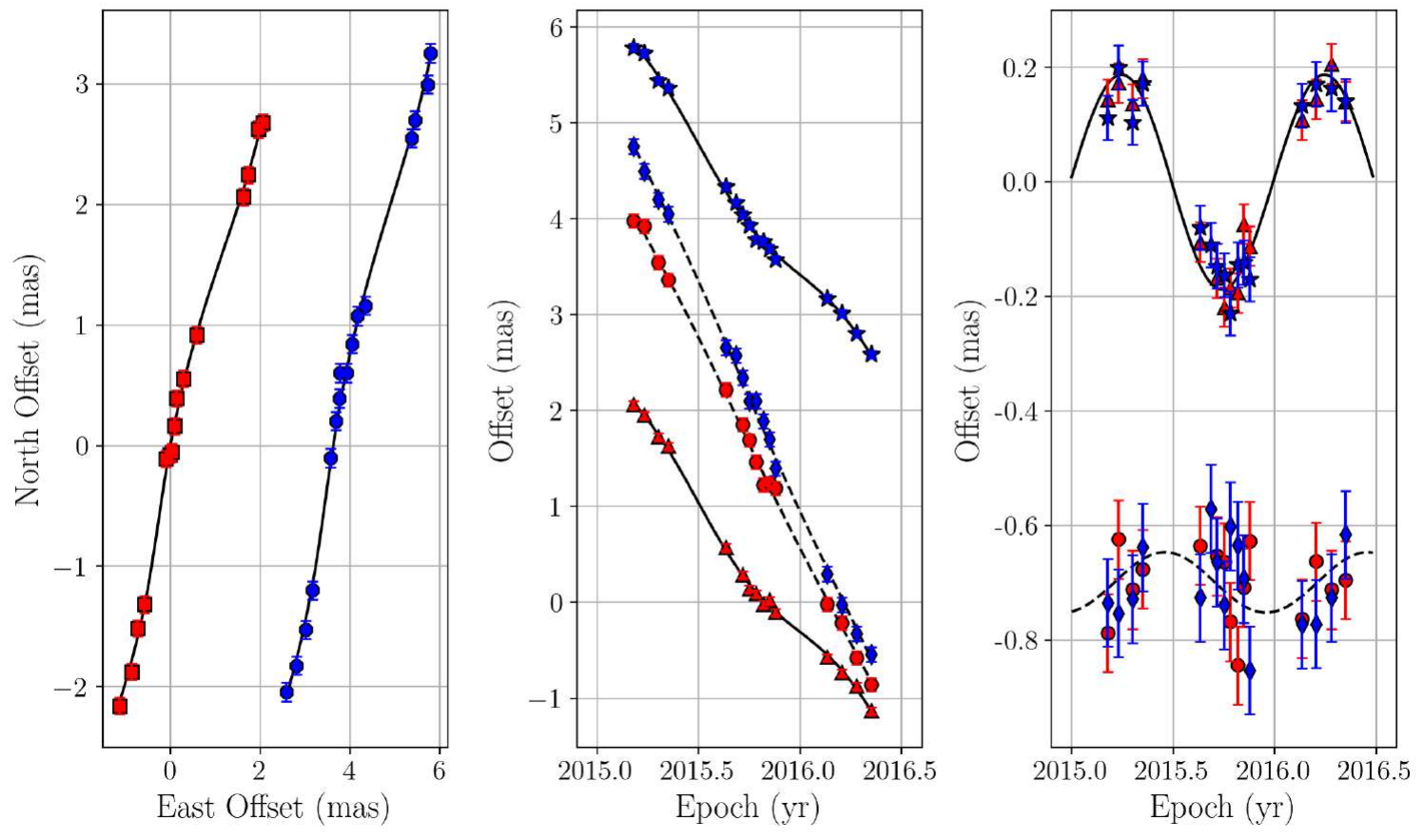}
\caption{Parallax and proper motion fits for the 110.68~km s$^{-1}$  maser spot of G024.78$+$00.08 with respect to calibrators J1846$-$0651 (red) \& J1825$-$0737 (blue). Different position components and calibrators are offset for clarity. \textit{Left Panel:} position offsets on the sky with respect to the two calibrators are shown by squares and octagons. \textit{Middle Panel:} position offsets and best fits in the easterly (triangles and stars and solid lines) and northerly (dots and diamonds and dashed lines) directions vs. time. \textit{Right Panel:} same as the middle panel, but with the fitted proper motion subtracted (i.e., displaying only the parallax effect).
}
\label{fig:appA6}
\end{figure}
\clearpage

\begin{figure}[htp]
\centering
\includegraphics[angle=0,scale=0.5]{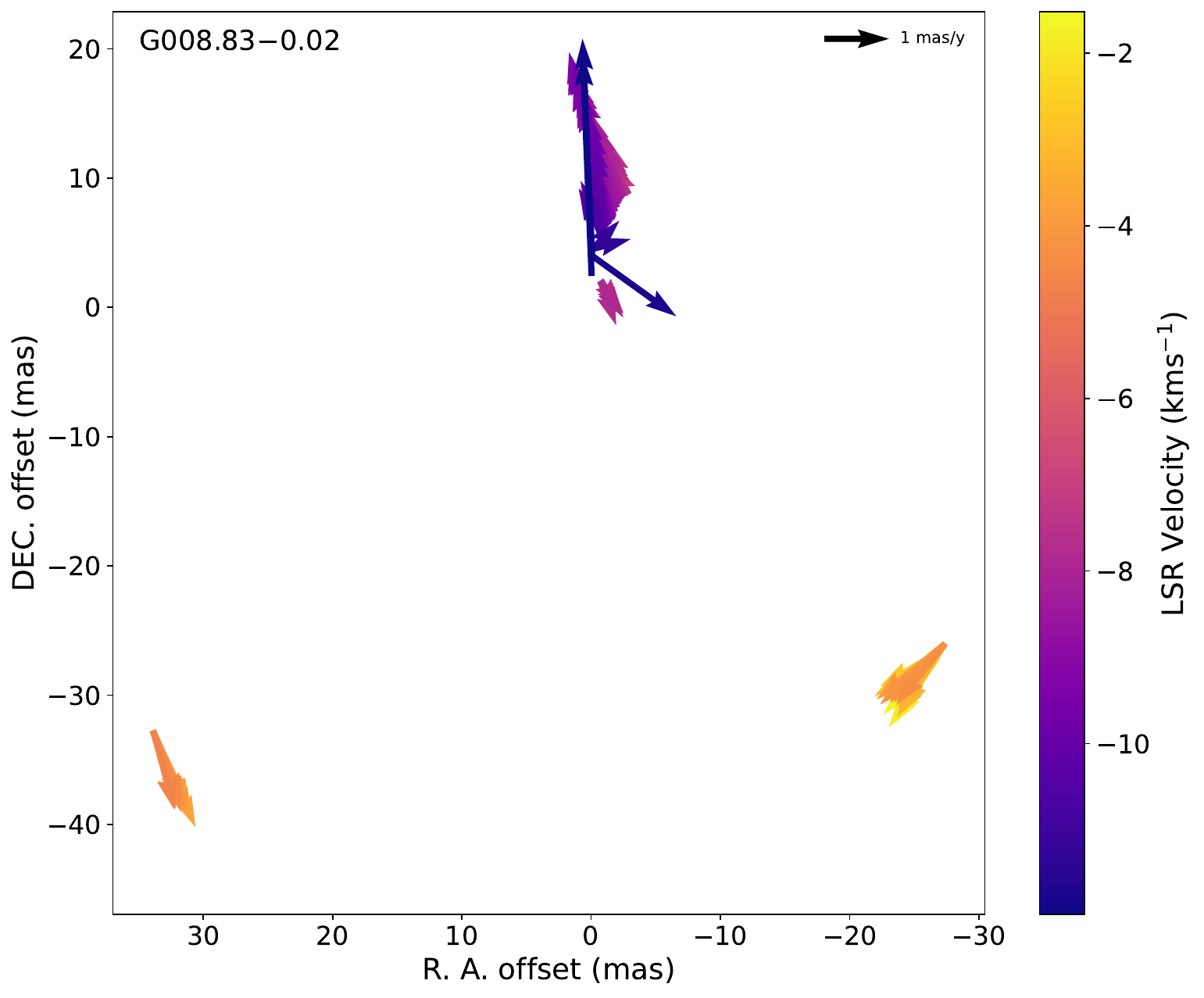}
\caption{Residual (internal) motions of maser spots in the water maser G008.83$-$00.02. The arrows represent the measured proper motions, with the average proper motion subtracted. The color of the arrows corresponds to the LSR velocity in ~km s$^{-1}$, as indicated by the color bar. The observed residual motions exhibit a systematic pattern, potentially indicating expansion or rotation within the region.
}
\label{fig:appA7}
\end{figure}
\clearpage

\begin{figure}[htp]
\centering
\includegraphics[angle=0,scale=0.5]{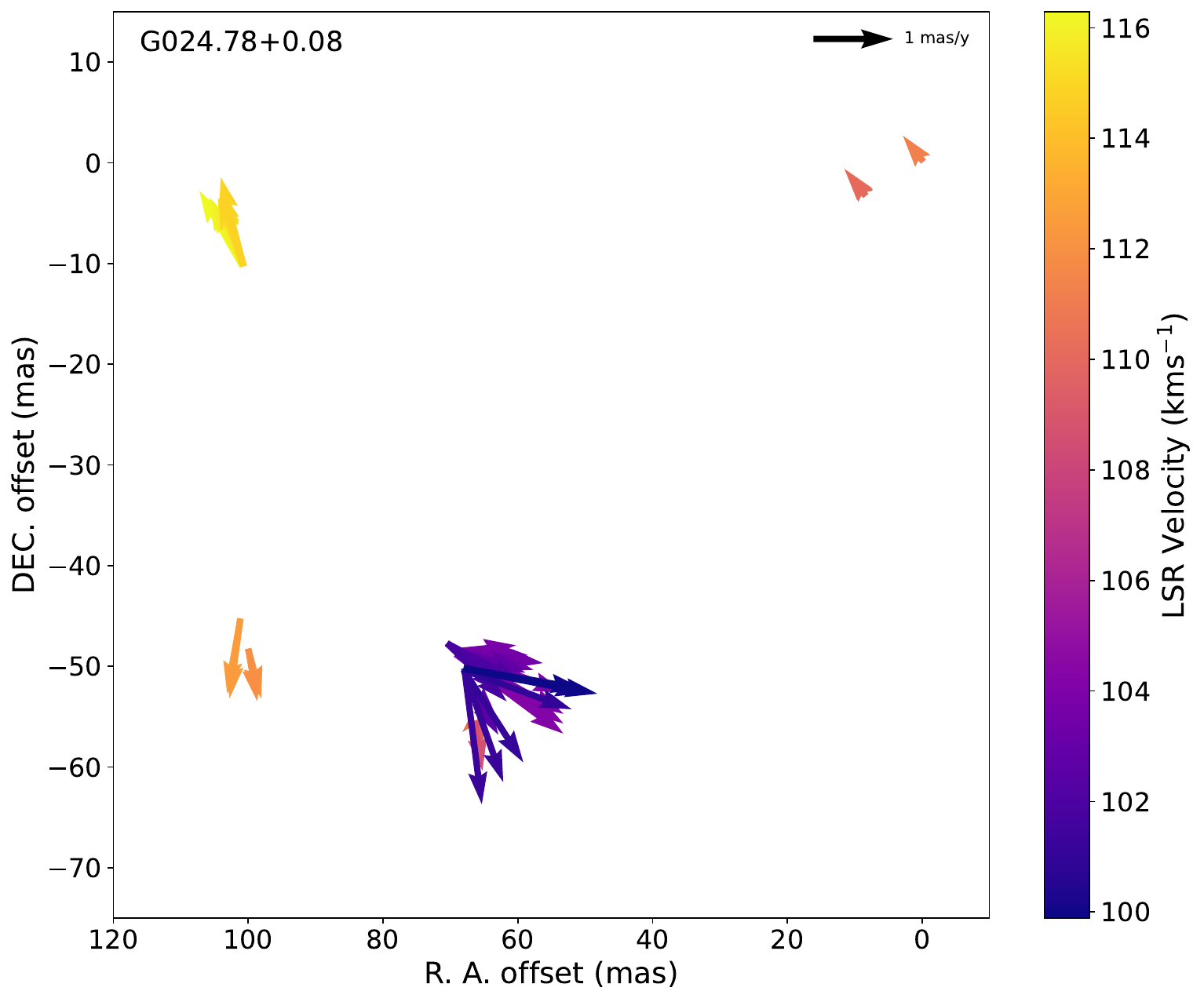}
\caption{Residual (internal) motions of maser spots in the water maser G024.78$+$00.08. Description same as in Figure~\ref{fig:appA7}.
}
\label{fig:appA8}
\end{figure}
\clearpage

\begin{figure}[htp]
\centering
\includegraphics[angle=0,scale=0.5]{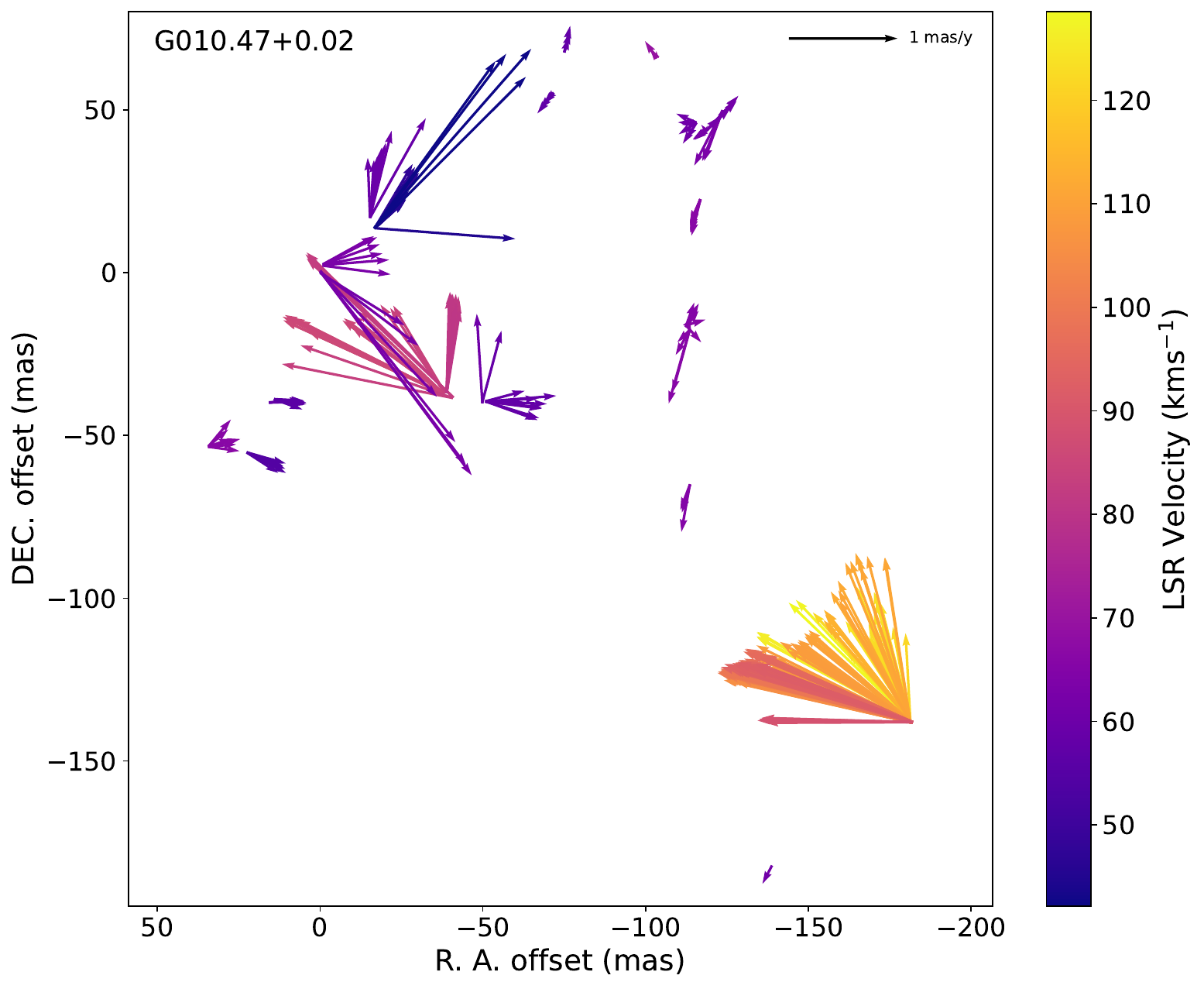}
\caption{Residual (internal) motions of maser spots in the water maser G010.47$+$00.02. Description same as in Figure~\ref{fig:appA7}.
}
\label{fig:appA9}
\end{figure}
\clearpage

\begin{figure}[htp]
\centering
\includegraphics[angle= 0, scale= 0.6]{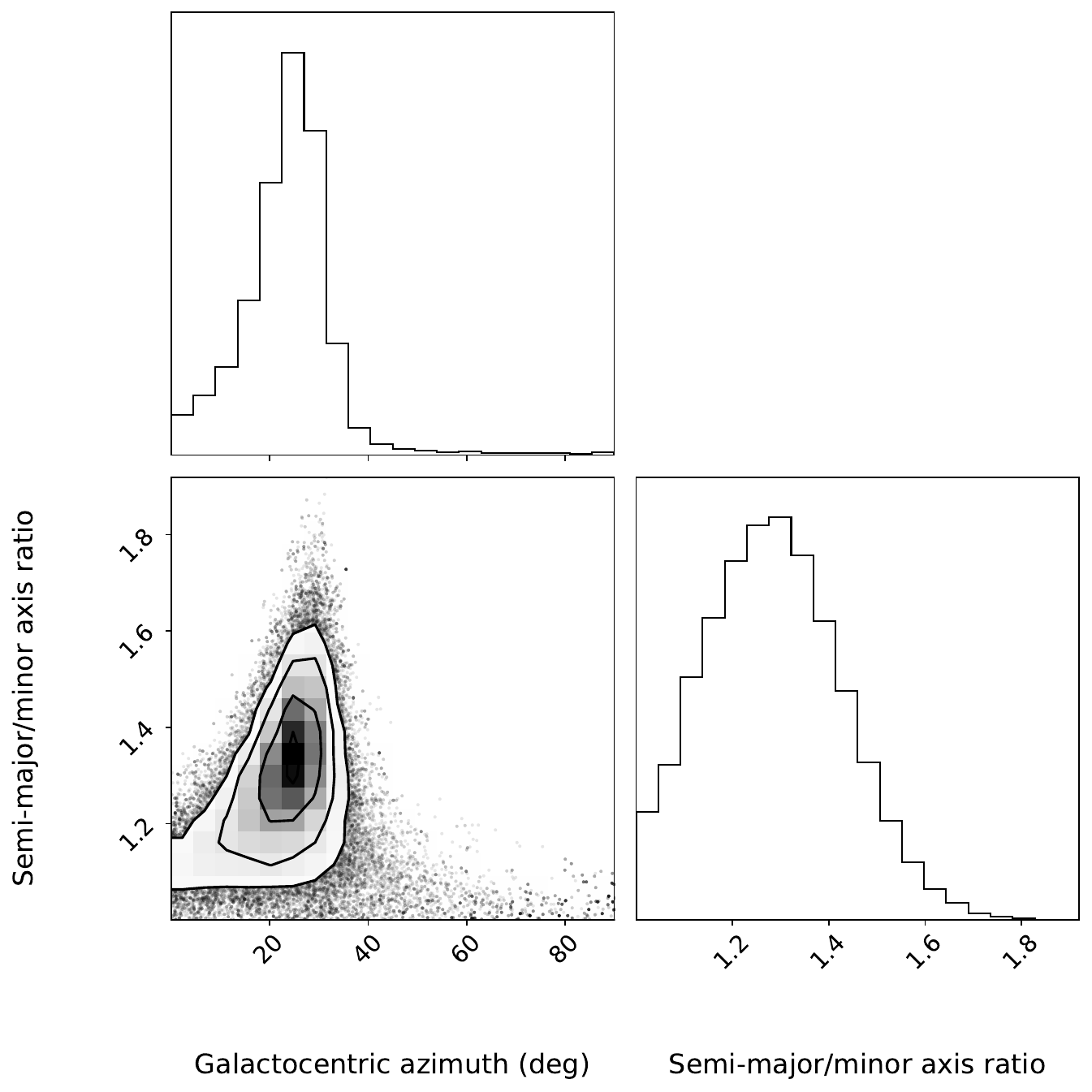}
\caption{MCMC Corner plot of the parameter distributions: The 2D marginal distributions for the Galactocentric azimuth (top left), semimajor/minor axis ratio (bottom right), and joint distributions between them (bottom left), showing the correlation between these parameters. The data points are derived from the MCMC sampling process described in Section~\ref{sec:ResultsDiscussion}.
}
\label{fig:appA10}
\end{figure}
\clearpage

\section{Tables}\label{sec:appendixB}

\renewcommand{\thetable}{B\arabic{table}} 
\setcounter{table}{0} 

\begin{table*}[htp]
\centering
\begin{tabular}{lccc} 
\toprule \toprule 
\multicolumn{1}{c}{\bf Epoch} &
\multicolumn{1}{c}{\bf G008.83$-$00.02} &
\multicolumn{1}{c}{\bf G010.47$+$00.02} &
\multicolumn{1}{c}{\bf G024.78$+$00.08} \\
\multicolumn{1}{c}{(1)} &
\multicolumn{1}{c}{(2)} &
\multicolumn{1}{c}{(3)} &
\multicolumn{1}{c}{(4)} \\
\toprule 
E1  &   2015 Mar 5   &  2015 Mar 5    &   2015 Mar 6     \\
E2  &   2015 Mar 22  &  2015 Mar 22   &   2015 Mar 25    \\
E3  &   2015 Apr 13  &  2015 Apr 13   &   2015 Apr 19    \\
E4  &   2015 May 1   &  2015 May 1    &   2015 May 8     \\
E5  &   2015 Aug 17  &  2015 Aug 17   &   2015 Aug 19    \\
E6  &   2015 Sep 1   &  2015 Sep 1    &   2015 Sep 7     \\
E7  &   2015 Sep 14  &  2015 Sep 14   &   2015 Sep 18    \\
E8  &   2015 Sep 29  &  2015 Sep 29   &   2015 Oct 1     \\
E9  &   2015 Oct 10  &  2015 Oct 10   &   2015 Oct 12    \\
E10 &   2015 Oct 25  &  2015 Oct 25   &   2015 Oct 26    \\
E11 &   2015 Nov 3   &  2015 Nov 3    &   2015 Nov 5     \\
E12 &   2015 Nov 15  &  2015 Nov 15   &   2015 Nov 17    \\
E13 &   2016 Feb 13  &  2016 Feb 13   &   2016 Feb 18    \\
E14 &   2016 Mar 8   &  2016 Mar 8    &   2016 Mar 15    \\
E15 &   2016 Apr 3   &  2016 Apr 3    &   2016 Apr 11    \\
E16 &   2016 Apr 29  &  2016 Apr 29   &   2016 May 7     \\
\bottomrule
\end{tabular}
\caption{Observation timeline for each epoch. Column (1): the epoch code. Columns (2)--(4) dates of observation.
}
\label{tab:appB1}
\end{table*}

\begin{table*}
\centering
\resizebox{\columnwidth}{!}{%
\begin{tabular}{lllrrrrrr} \toprule \toprule
\multicolumn{2}{c}{\bf Source Name} &
\multicolumn{1}{c}{\bf R.A.(J2000)}  &
\multicolumn{1}{c}{\bf Dec.(J2000)}  &
\multicolumn{1}{c}{\bf Separation}   &
\multicolumn{1}{c}{\bf P.A.}   &
\multicolumn{1}{c}{\bf Brightness}   &
\multicolumn{1}{c}{$ \bf V_{\rm \bf LSR}$}   & 
\multicolumn{1}{c}{\bf Beam}   \\
\multicolumn{1}{c}{\bf Maser} &
\multicolumn{1}{c}{\bf Calibrators} &
\multicolumn{1}{c}{\bf (h m s)} &
\multicolumn{1}{c}{\bf (deg, arcmin, arcsec)} &
\multicolumn{1}{c}{\bf $\theta_\mathrm{sep}$(deg)}  &
\multicolumn{1}{c}{\bf (deg)} &
\multicolumn{1}{c}{\bf (Jy~Beam$^{-1}$)} &
\multicolumn{1}{c}{\bf ({km~s$^{-1}$})}   &
\multicolumn{1}{c}{\bf (mas, mas, deg)}   \\
\multicolumn{1}{c}{(1)} &
\multicolumn{1}{c}{(2)} &
\multicolumn{1}{c}{(3)} &
\multicolumn{1}{c}{(4)} &
\multicolumn{1}{c}{(5)} &
\multicolumn{1}{c}{(6)} &
\multicolumn{1}{c}{(7)} &
\multicolumn{1}{c}{(8)} &
\multicolumn{1}{c}{(9)} \\
\hline\hline
G008.83$-$00.02   &               & 18~05~25.6927  & $-$21~19~25.397  &     &        & 07.78 & $-$12.74 & 2.10 $\times$ 0.70 @ \phantom{0}$-$2  \\
                  & J1813$-$2015  & 18~13~16.8731  & $-$20~15~44.064  & 2.1 & 60     & 0.006 &          & 2.14 $\times$ 0.64 @ \phantom{0}$-$4  \\
                  & J1755$-$2232  & 17~55~26.2848  & $-$22~32~10.617  & 2.6 & $-$117 & 0.062 &          & 2.27 $\times$ 0.67 @ \phantom{0}$-$1  \\
\hline
G010.47$+$00.02   &               & 18~08~38.2338  & $-$19~51~50.303  &     &        & 36.95 & 61.68    & 2.05 $\times$ 0.70 @ \phantom{0}$-$2  \\
                  & J1808$-$1822  & 18~08~55.5155  & $-$18~22~53.390  & 1.5 & 3      & 0.011 &          & 2.00 $\times$ 0.69 @ \phantom{0}$-$2  \\
                  & J1813$-$2015  & 18~13~16.8731  & $-$20~15~44.064  & 1.2 & 110    & 0.006 &          & 2.05 $\times$ 0.71 @ \phantom{0}$-$1  \\
\hline
G024.78$+$00.08   &               & 18~36~12.5490  & $-$07~12~10.780  &     &        & 14.84 & 110.68   & 1.51 $\times$ 0.54 @ $-$12 \\
                  & J1846$-$0651  & 18~46~06.3003  & $-$06~51~27.746  & 2.5 & 83     & 0.029 &          & 1.60 $\times$ 0.67 @ \phantom{0}$-$5  \\
                  & J1825$-$0737  & 18~25~37.6096  & $-$07~37~30.013  & 2.7 & $-$100 & 0.084 &          & 1.80 $\times$ 0.70 @ \phantom{0}$-$2  \\
\hline
\bottomrule \bottomrule
\end{tabular}%
}
\caption{The observational parameters and other information for each observed source (masers and calibrators). Columns (1) and (2): the maser and calibrator names. Columns (3) and (4): R.A. and decl. Columns (5) and (6): the angular separation for a calibrator along a position angle (P.A.) east of north from the maser. Columns (7)–-(9): the peak apparent brightness, LSR velocity (for the maser reference channel), and size and position angle of the uniformly weighted CLEAN beam. All values are for the reference epochs: E10 for G008.83$-$00.02, E10 for G010.47$+$00.02, and E9 for G024.78$+$00.08.
}
\label{tab:appB2}
\end{table*}
\clearpage

\begin{table*}[htb]
\centering
\resizebox{\columnwidth}{!}{%
\begin{tabular}{lrrrrrrr} \toprule \toprule
\multicolumn{1}{c}{\bf Background} & 
\multicolumn{1}{c}{\bf $ \bf V_{\bf LSR}$}      & 
\multicolumn{1}{c}{\bf Detected}   & 
\multicolumn{1}{c}{\bf Parallax}   & 
\multicolumn{1}{c}{\bf $\mu_x$ }   & 
\multicolumn{1}{c}{\bf $\mu_y$}    & 
\multicolumn{1}{c}{\bf $ \bf \Delta x$}  & 
\multicolumn{1}{c}{\bf $\bf \Delta y$} \\
\multicolumn{1}{c}{\bf Calibrator}  & 
\multicolumn{1}{c}{\bf (km~s$^{-1}$)}  & 
\multicolumn{1}{c}{\bf Epochs}  & 
\multicolumn{1}{c}{\bf (mas)}   & 
\multicolumn{1}{c}{\bf (mas~yr$^{-1}$)} & 
\multicolumn{1}{c}{\bf (mas~yr$^{-1}$)} & 
\multicolumn{1}{c}{\bf (mas)}   & 
\multicolumn{1}{c}{\bf (mas)}   \\
\multicolumn{1}{c}{(1)} &
\multicolumn{1}{c}{(2)} &
\multicolumn{1}{c}{(3)} &
\multicolumn{1}{c}{(4)} &
\multicolumn{1}{c}{(5)} &
\multicolumn{1}{c}{(6)} &
\multicolumn{1}{c}{(7)} &
\multicolumn{1}{c}{(8)} \\
\hline\hline
~~~\\
\multicolumn{8}{c}{\bf G008.83$-$00.02}\\\hline
J1755$-$2232 & $-$12.74 & 1111 11110111 1111 &  0.185$\pm$0.019 &      $-$1.18$\pm$ 0.05 &   $-$1.43$\pm$ 0.21 &   0.446$\pm$      0.017 &       0.220$\pm$      0.077 \\
& $-$3.46  & 1111 11110111 1111 &  0.186$\pm$0.020 &      $-$1.97$\pm$ 0.05 &   $-$2.23$\pm$ 0.22 &   $-$26.845$\pm$  0.017 &   $-$25.786$\pm$      0.080 \\
& $-$4.22  & 1111 11110111 1100 &  0.220$\pm$0.023 &      $-$0.69$\pm$ 0.07 &   $-$2.30$\pm$ 0.26 &   34.353$\pm$     0.019 &   $-$32.382$\pm$      0.084 \\
& $-$10.36 & 1111 11110111 1111 &  0.242$\pm$0.016 &      $-$1.58$\pm$ 0.04 &   $-$0.46$\pm$ 0.30 &   $-$0.521$\pm$   0.014 &       5.617$\pm$      0.108 \\
\multicolumn{3}{c}{Combined fit} & 0.206$\pm$0.020 &    &  \\
\hline
J1813$-$2015 & $-$12.74  & 1111 00110110 1111 &  0.212$\pm$0.076 &      $-$1.07$\pm$ 0.16 &   $-$1.04$\pm$ 0.19 &   0.156$\pm$     0.068 &   0.048$\pm$          0.077 \\
& $-$3.46  & 1111 00110110 1111 &  0.221$\pm$0.078 &      $-$1.84$\pm$ 0.16 &   $-$1.82$\pm$ 0.22 &   $-$27.139$\pm$ 0.070 &   $-$25.973$\pm$      0.087 \\
& $-$4.22  & 1111 00110110 1100 &  0.217$\pm$0.106 &      $-$0.53$\pm$ 0.27 &   $-$2.23$\pm$ 0.25 &   34.036$\pm$    0.093 &   $-$32.649$\pm$      0.090 \\
& $-$10.36 & 1111 00110110 1111 &  0.277$\pm$0.085 &      $-$1.45$\pm$ 0.18 &   $-$0.07$\pm$ 0.25 &   $-$0.813$\pm$  0.077 &      5.412$\pm$       0.101 \\
\multicolumn{3}{c}{Combined fit} & 0.233$\pm$0.082 &   &  \\
\hline
Average   &&&0.208$\pm$0.019 & $-$1.35$\pm$ 0.27  & $-$1.43$\pm$ 0.46 & \\
\hline\hline
~~~\\
\multicolumn{8}{c}{\bf G010.47$+$00.02}\\\hline
J1808$-$1822 & 61.68 & 1111 01111111 1111 &     0.091$\pm$0.029 &      $-$2.96$\pm$ 0.07 &      $-$6.45$\pm$ 0.14 &        0.216$\pm$     0.025 &       0.222$\pm$      0.052 \\
& 46.15 & 1111 01111111 1111 &     0.098$\pm$0.029 &      $-$3.46$\pm$ 0.07 &      $-$5.83$\pm$ 0.14 &     $-$16.556$\pm$    0.026 &      13.887$\pm$      0.052 \\
& 56.61 & 1111 01111111 1111 &     0.098$\pm$0.029 &      $-$3.46$\pm$ 0.07 &      $-$6.09$\pm$ 0.14 &        14.376$\pm$    0.026 &   $-$38.679$\pm$      0.052 \\
& 66.75 & 1111 01111111 1111 &     0.095$\pm$0.028 &      $-$3.83$\pm$ 0.07 &      $-$6.31$\pm$ 0.14 &    $-$116.588$\pm$    0.025 &      22.748$\pm$      0.052 \\
& 54.56 & 1111 01111111 1110 &     0.101$\pm$0.034 &      $-$3.35$\pm$ 0.09 &      $-$6.16$\pm$ 0.16 &        22.805$\pm$    0.030 &   $-$54.983$\pm$      0.056 \\
& 64.92 & 1111 01111111 1111 &     0.093$\pm$0.028 &      $-$3.82$\pm$ 0.07 &      $-$6.31$\pm$ 0.15 &    $-$113.399$\pm$    0.025 &   $-$64.806$\pm$      0.053 \\
& 65.78 & 1111 01111111 1111 &     0.088$\pm$0.027 &      $-$3.86$\pm$ 0.07 &      $-$6.30$\pm$ 0.14 &    $-$113.170$\pm$    0.024 &   $-$17.611$\pm$      0.052 \\
& 62.44 & 1111 01111111 1110 &     0.105$\pm$0.030 &      $-$3.80$\pm$ 0.08 &      $-$6.16$\pm$ 0.16 &    $-$118.278$\pm$    0.026 &      43.939$\pm$      0.056 \\
\multicolumn{3}{c}{Combined fit} & 0.096$\pm$0.028  &   &  \\
\hline
J1813$-$2015  & 61.68 & 1111 01010111 1111 &     0.102$\pm$0.036 &     $-$2.84$\pm$ 0.08 &     $-$6.52$\pm$ 0.09 &      $-$3.623$\pm$     0.031 &       5.577$\pm$        0.038 \\
& 46.15 & 1111 01010111 1111 &     0.113$\pm$0.035 &     $-$3.34$\pm$ 0.07 &     $-$5.90$\pm$ 0.10 &     $-$20.403$\pm$     0.031 &      19.249$\pm$        0.039 \\
& 56.61 & 1111 01010111 1111 &     0.107$\pm$0.036 &     $-$3.35$\pm$ 0.08 &     $-$6.18$\pm$ 0.09 &        10.531$\pm$     0.031 &   $-$33.322$\pm$        0.037 \\
& 66.75 & 1111 01010111 1111 &     0.106$\pm$0.037 &     $-$3.71$\pm$ 0.08 &     $-$6.38$\pm$ 0.09 &    $-$120.433$\pm$     0.032 &      28.113$\pm$        0.039 \\
& 54.56 & 1111 01010111 1110 &     0.096$\pm$0.041 &     $-$3.30$\pm$ 0.09 &     $-$6.29$\pm$ 0.08 &        18.948$\pm$     0.035 &   $-$49.639$\pm$        0.034 \\
& 64.92 & 1111 01010111 1111 &     0.101$\pm$0.035 &     $-$3.71$\pm$ 0.07 &     $-$6.39$\pm$ 0.09 &    $-$117.241$\pm$     0.031 &   $-$59.436$\pm$        0.039 \\
& 65.78 & 1111 01010111 1111 &     0.098$\pm$0.035 &     $-$3.74$\pm$ 0.07 &     $-$6.36$\pm$ 0.13 &    $-$117.013$\pm$     0.031 &   $-$12.263$\pm$        0.051 \\
& 62.44 & 1111 01010111 1110 &     0.103$\pm$0.037 &     $-$3.75$\pm$ 0.08 &     $-$6.30$\pm$ 0.09 &    $-$122.136$\pm$     0.032 &      49.284$\pm$        0.037 \\
\multicolumn{3}{c}{Combined fit} & 0.103$\pm$0.035  &   &  \\
\hline
Average   &&&0.099$\pm$0.022 & $-$3.52$\pm$ 0.11  & $-$6.27$\pm$ 0.07 & \\
\hline\hline
~~~\\
\multicolumn{8}{c}{\bf G024.78$+$00.08}\\\hline
J1846$-$0651 & 110.68 & 1111 10111111 1111 &     0.175$\pm$0.010 &      $-$2.72$\pm$ 0.02 &      $-$4.21$\pm$ 0.05 &      0.361$\pm$       0.009 &    0.357$\pm$      0.017 \\
& 115.53 & 0111 10111111 1111 &     0.205$\pm$0.011 &      $-$2.80$\pm$ 0.03 &      $-$3.81$\pm$ 0.05 &     101.132$\pm$      0.010 &    $-$9.858$\pm$   0.018 \\
& 111.86 & 1111 10111111 1111 &     0.202$\pm$0.011 &      $-$2.24$\pm$ 0.03 &      $-$5.17$\pm$ 0.08 &     100.289$\pm$      0.009 &    $-$47.937$\pm$  0.029 \\
& 110.03 & 1111 10111111 1111 &     0.184$\pm$0.008 &      $-$2.58$\pm$ 0.02 &      $-$4.28$\pm$ 0.04 &     8.778$\pm$        0.007 &    $-$2.908$\pm$   0.015 \\
\multicolumn{3}{c}{Combined fit} &  0.191$\pm$0.010 &   &  \\
\hline
J1825$-$0737 & 110.68 & 1111 11111111 1111 &     0.168$\pm$0.011 &      $-$2.75$\pm$ 0.03 &      $-$4.61$\pm$ 0.06 &      4.095$\pm$       0.010 &    0.642$\pm$       0.019 \\
& 115.53 & 0111 11111111 1111 &     0.202$\pm$0.011 &      $-$2.84$\pm$ 0.03 &      $-$4.22$\pm$ 0.06 &     104.868$\pm$      0.010 &    $-$9.575$\pm$    0.019 \\
& 111.86 & 1111 11111111 1111 &     0.198$\pm$0.010 &      $-$2.27$\pm$ 0.02 &      $-$5.58$\pm$ 0.07 &     104.019$\pm$      0.009 &    $-$47.661$\pm$   0.026 \\
& 110.03 & 1111 11111111 1111 &     0.176$\pm$0.010 &      $-$2.61$\pm$ 0.02 &      $-$4.70$\pm$ 0.05 &     12.511$\pm$       0.009 &    $-$2.626$\pm$    0.017 \\
\multicolumn{3}{c}{Combined fit} &  0.185$\pm$0.010 &   &  \\
\hline
Average   &&&0.188$\pm$0.007 & $-$2.60$\pm$ 0.12  & $-$4.55$\pm$ 0.30 & \\
\hline\hline
\bottomrule
\end{tabular}%
}
\caption{Observational results for the three maser targets. Column (1): the background calibrator name. Column (2): the LSR velocity of a maser spot. Column (3): the detection of a maser spot and calibrator by epoch: 1 means detected and 0 not detected. Columns (4), (5), and (6): the fitted parallax and proper motions along the easterly and northerly directions. Columns (7) and (8): the position offset from the reference maser channel listed in Table 3. In the ``Combined fit,'' we performed a parallax fit with the listed maser spots, weighting the data by combining in quadrature individual-fit error floors with the formal measurement uncertainties. The parallax uncertainty has been inflated by a factor of $\sqrt{N}$, where $N$ is the number of spots.  For the ``Average'' of parallaxes with both calibrators, we used a variance-weighted average of the combined fits; for the average proper motion, we used a hybrid approach by first doing a variance-weighted average of proper motion components for both calibrators for each maser spot and then an unweighted average of all the maser spots; uncertainties are standard errors of the mean, based on the scatter in the data.
}
\label{tab:appB3}
\end{table*}
\clearpage

\begin{table*}[htb]
\centering
\resizebox{\columnwidth}{!}{%
\begin{tabular}{lrrrrr} \toprule \toprule
\multicolumn{1}{c}{\bf Target} &
\multicolumn{1}{c}{\bf Parallax}  &
\multicolumn{1}{c}{\bf Distance}  &
\multicolumn{2}{c}{\bf Proper Motion}   &
\multicolumn{1}{c}{\bf LSR Velocity}   \\
\multicolumn{1}{c}{\bf Maser} &
\multicolumn{1}{c}{\bf $\bf \pi$~(mas)}     &
\multicolumn{1}{c}{\bf  D (kpc)}  &
\multicolumn{1}{c}{\bf $\bf \mu_x$({mas~y$^{-1}$})}&
\multicolumn{1}{c}{\bf $\bf \mu_y$({mas~y$^{-1}$})}&
\multicolumn{1}{c}{\bf (km~s$^{-1}$)}\\
\multicolumn{1}{c}{(1)} &
\multicolumn{1}{c}{(2)} &
\multicolumn{1}{c}{(3)} &
\multicolumn{1}{c}{(4)} &
\multicolumn{1}{c}{(5)} &
\multicolumn{1}{c}{(6)} \\
\hline
\hline
G008.83$-$00.02    & $0.208\pm0.019$ &  $4.81^{+ 0.48}_{- 0.40}$ & $-1.35\pm0.27$  & $-1.43\pm0.46$ &  $1^{a}\pm 5$ \\
G010.47$+$00.02    & $0.115\pm0.008$ &  $8.70^{+0.65}_{-0.57}$   & $-3.69\pm0.08$  & $-6.34\pm0.07$ &  $69^{b}\pm 5$ \\
G024.78$+$00.08    & $0.188\pm0.007$ &  $5.32^{+ 0.20}_{- 0.19}$ & $-2.60\pm0.12$  & $-4.55\pm0.30$ &  $110^{c}\pm 5$ \\
\hline
\bottomrule \bottomrule
\end{tabular}%
}
\caption{Final Results for Parallax and Proper Motion. Column (1): the Galactic source name/coordinates. Columns (2)–(5): the parallax, the implied parallax distance, and proper motion in the eastward ($\mu_x$ = $\mu_\alpha\cos\delta$) and northward ($\mu_y$ =$\mu_\delta$ ) directions. Column (6): the LSR velocity from molecular line emission. $V_{\rm LSR}$ references: (a) NH$_{3}$ (1,1) \citep{Molinari:1996}; (b) CS(2$-$1) \citep{Bronfman:1996} ; (c) CS(1$-$0) \citep{Anglada:1996}.
}
\label{tab:appB4}
\end{table*}

\section{X1 Orbit Model -- Calculation of Tangential Velocity}\label{sec:appendixC}

This section provides details of the X1 orbit model. \((x, y)\) are the coordinates of the maser source in the rotated frame, while \((x_{\textit{unrot}}, y_{\textit{unrot}})\) are its coordinates in the unrotated frame. \(\theta\) represents the rotation angle of the ellipse measured counterclockwise from the positive x-axis, and \textit{axis\_ratio} is the ratio of the semimajor axis (\(a\)) to the semiminor axis (\(b\)) of the X1 orbit. The Galactocentric azimuth is given by \(90^\circ - \theta\). \((T_{\text{x}}, T_{\text{y}})\) denote the components of the tangential vector in the unrotated frame, while \((T_{\text{v\_x}}, T_{\text{v\_y}})\) are the tangential velocity components in the rotated frame. \textit{S} represents the circular rotation velocity at a Galactocentric radius \(R\) to the maser.  

\begin{equation}
x_{\text{unrot}} = x \cos(\theta) + y \sin(\theta)
\end{equation}

\begin{equation}
y_{\text{unrot}} = -x \sin(\theta) + y \cos(\theta)
\end{equation}

\begin{equation}
\text{b} = \sqrt{\left( \frac{x_{\text{unrot}}}{\text{axis ratio}} \right)^2 + y_{\text{unrot}}^2}
\end{equation}

\begin{equation}
\text{a} = \text{axis ratio} \times \text{b}
\end{equation}

\begin{equation}
\text{radial angle} = \arctan\left( \frac{y_{\text{unrot}}}{b} \Big/ \frac{x_{\text{unrot}}}{a} \right)
\end{equation}

\begin{equation}
T_{\text{x}} = \text{a} \sin(\text{radial angle})
\end{equation}

\begin{equation}
T_{\text{y}} = -\text{b} \cos(\text{radial angle})
\end{equation}

\begin{equation}
T_{\text{v\_x}} = \text{S} \left( \hat{\bm{T}}_{\text{x}}  \cos(\theta) - \hat{\bm{T}}_{\text{y}}  \sin(\theta) \right)
\end{equation}

\begin{equation}
T_{\text{v\_y}} = \text{S} \left( \hat{\bm{T}}_{\text{x}}  \sin(\theta) + \hat{\bm{T}}_{\text{y}}  \cos(\theta) \right)
\end{equation}

\clearpage


\bibliography{JayReference}{}
\bibliographystyle{aasjournal}



\end{document}